\documentclass[useAMS,usenatbib,usegraphicx]{mn2e}
\usepackage{bm}

\begin{document}
%Issue 2.4
\title[Made-to-measure galaxy models]{Made-to-measure galaxy models - I Methodology}
\author[R. J. Long and S. Mao]
  {R. J.~Long$^1$\thanks{E-mail: rlong@jb.man.ac.uk} and S.~Mao$^1$\thanks{E-mail: shude.mao@manchester.ac.uk} \\
   $^1$Jodrell Bank Centre for Astrophysics, Alan Turing Building, The University of Manchester, Manchester M13 9PL\\}
\date{Accepted 2010 January 27.  Received 2010 January 26; in original form 2009 February 27}
\pagerange{\pageref{firstpage}--\pageref{lastpage}} \pubyear{2010}
\maketitle 

\label{firstpage}

\begin{abstract}
We re-derive the made-to-measure method of \citet{Syer1996} for modelling stellar systems and individual galaxies, and demonstrate how extensions to the made-to-measure method may be implemented and used.  We illustrate the enhanced made-to-measure method by determining the mass-to-light ratio of a galaxy modelled as a Plummer sphere.   From the standard galactic observables of surface brightness and line-of-sight velocity dispersion together with the $h_4$ Gauss-Hermite coefficient of the line-of-sight velocity distribution,  we successfully recover the true mass-to-light ratio of our toy galaxy.  Using kinematic data from \citet{Kleyna2002}, we then estimate the mass-to-light ratio of the dwarf spheroidal galaxy Draco achieving a V-band value of $539\pm 136 \ \rmn{M}_{\odot}/\rmn{L}_{\odot}$.  We describe the main aspects of creating a made-to-measure galaxy model and show how the key modelling parameters may be determined.
\end{abstract}

\section{Introduction}\label{sec:intro}
Within the field of galactic and stellar dynamics, it has become common practice to model kinematic observations of a galaxy in order to interpret the observations and to understand better the underlying dynamical structures within the galaxy.  N-body modelling is one of the techniques employed.  \citet{Syer1996} categorised methods for creating N-body systems into 3 groups namely distribution function based, moment or Jeans equation based and orbit based, placing their made-to-measure (M2M) method in a new particle based group. The M2M method is, however, similar to the methods in the orbit based group notably the method of \citet{Schwarz1979}.  Schwarzschild's method since its inception has undergone both significant development (for example, \citealt{Chaname2008}) and exploitation (for example,  \citealt{SauronX2007}). \citet{Chaname2008} extended the method to handle discrete stellar kinematic data sets, and \citet{SauronX2007} used Schwarzschild's method to model photometric and kinematic observations of elliptical galaxies from the SAURON survey.  By comparison, Syer and Tremaine's M2M method remained largely unutilised until the bulge and disk model of the Milky Way in \citet{Bissantz2004}.  More recently, the M2M method was enhanced by the inclusion of a capability to model kinematic data (\citealt{EJ2007}, \citealt{DL2007} and \citealt{DL2008a}), and was applied to model the elliptical galaxies NGC3377 and NGC4697.  More recently still, \citet{Dehnen2009} implemented an alternative weight adaption mechanism within the M2M method.

In this paper we remain true to the outline in \citet{Syer1996} but reframe the M2M method (section \ref{sec:m2m}) slightly to improve the theoretical basis for the weight evolution equation, and introduce 2 further constraints - the first on the sum of the particle weights and the second on the isotropy of the velocity dispersion.  We show how the parameters necessary to tune the model for a given situation may be determined (section \ref {sec:basicmodelcap}).  We use the method to determine the mass-to-light ratio of a toy galaxy (section \ref {sec:appml}) and then to estimate the mass-to-light ratio of the dwarf spheroidal galaxy Draco (section \ref{sec:dracomain}). We aim to provide sufficient detail and advice to enable others to produce their own implementation of the M2M method. Our implementation is described in section \ref{sec:implementation}.

\section{The M2M Method}\label{sec:m2m}
\subsection{Outline}
In brief, the M2M method is concerned with modelling stellar systems and individual galaxies as a system of test particles orbiting in a gravitational potential.  Weights are associated with the particles and are evolved over time (many orbital periods) such that, by using these weights, observational measurements of a real galaxy are reproduced.  The method uses these observational measurements as constraints on the model. Whilst it is tempting to think of the particles as representing stars, the particles are more accurately described as phase space density elements \citep{Hernquist1992} whose motion is integrated along the characteristic curves of the collisionless Boltzmann equation.  The gravitational potential may be prespecified or determined self-consistently, and may contain a dark matter component. \citet{Schwarz1979} uses a similar approach but does not modify the weights during the main modelling run obtaining them instead at the end via linear programming.

Reviewed in the next section, section \ref{sec:theory}, are the theoretical approaches taken in \citet{Syer1996} and \citet{DL2007}. We then recast the method from a maximum likelihood starting point paying particular attention to the origin of the derivative term in the weight evolution equation.  The particle weight convergence analysis in \citet{Syer1996} is not revisited in this paper.

\subsection{Theory}\label{sec:theory}
The galactic observables used with a M2M model are moments of the distribution function and have the general form
\begin{equation}
	Y_j = \int K_j(\bmath{r},\bmath{v}) f(\bmath{r},\bmath{v}) d^3\bmath{r}d^3\bmath{v},
\end{equation}
where subscript $j$ indicates an instance of the observable, $K_j$ is the kernel for the observation $Y_j$ and $f(\bmath{r},\bmath{v})$ is the phase space distribution function.  In this context, typical observables are surface brightness and surface brightness times the luminosity weighted line-of-sight velocity dispersion squared. For a model of $N$ particles, this integral form is translated into
\begin{equation}
	y_j(t) = \sum _i^N w_i(t)K_j(\bmath{r}_i (t), \bmath{v}_i (t)),
\end{equation}
where index $i$ runs from $1$ to $N$.  Note that $K_j$ embodies a selection function such that only the particles which have a direct effect on the observation $y_j$ are included in the sum.  The goal of the M2M method is to evolve the particle weights, $w_i(t)$, such that the time averaged model observations $y_j$ match the actual observations $Y_j$.

The weights, representing (in our case) the luminosity of individual particles implemented as a fraction of the total luminosity of the galaxy being modelled, are evolved using
\begin{equation}
	\frac{d}{dt}w_i(t)=-\epsilon w_i(t) \sum _j^J \frac{K_j(\bmath{r}_i(t),\bmath{v}_i(t))}{Z_j}\Delta_j(t),
\label{eqn:wtsyer}
\end{equation}
where index $j$ runs from $1$ to $J$, $\Delta _j(t) = \frac{y_j(t)-Y_j}{Y_j}$, $Z_j$ is arbitrary, and the kernels $K_j$ are implemented by binning the model's particle data.  $\epsilon$ is small, positive and constant.  Calculations of the $\Delta _j$ from the model are exponentially smoothed to reduce the impact of particle counting effects using
\begin{equation}
	\frac{d}{dt}\tilde{\Delta}_j(t) = \alpha (\Delta _j(t) - \tilde{\Delta}_j(t)),
\label{eq:tempsmooth}
\end{equation}
where $\alpha$ is small, positive and constant.  $\Delta _j(t)$ is then replaced with $\tilde{\Delta}_j(t)$ in equation \ref{eqn:wtsyer}.  \citet{DL2007} take the equivalent approach of smoothing $y_j$. 

The weight evolution equation may be extended in 2 ways - firstly by the introduction of a profit function \citep{Syer1996} to constrain the overall weight evolution and smooth observable reproduction (regularisation), and secondly by the inclusion of observational errors \citep{DL2007}.  The equation then becomes
\begin{equation}
	\frac{dw_i}{dt} = \epsilon w_i \frac{\partial F}{\partial w_i},
\label{eq:wtevolve}
\end{equation}
where 
\begin{equation}
	F = \mu S - \frac{1}{2} \chi ^2
\label{eqn:firstf}
\end{equation}
and is to be maximised.  $\epsilon$, $\alpha$ and $\mu$ in equations \ref{eqn:wtsyer}, \ref{eq:tempsmooth} and \ref{eqn:firstf} are discussed in more detail in \citet{Syer1996}. Here, the equivalent discussion is delayed until  section \ref{sec:paramtuning}.  Note that to arrive at equation \ref{eqn:wtsyer} from equation \ref{eqn:firstf} the kernels must not depend on the particle weights.  $S$, the profit function, varyingly known as the relative entropy or Kullback-Leibler divergence from some initial value (prior) of the particle weights, is given by
\begin{equation}
	S = - \sum _i^N w_i \ln \left( \frac{w_i}{m_i} \right),
\end{equation}
where the $m_i$ are the priors.  $\chi ^2$ is calculated as
\begin{equation}
	\chi ^2 = \sum_j^J \tilde{\Delta} _j^2,
\label{eq:chi2}
\end{equation}
with $\tilde{\Delta} _j$ being the exponentially smoothed form of either a relative difference between the model and target observations
\begin{equation}
	\Delta_j = \frac{y_j - Y_j}{Y_j},
\label{eq:withouterror}
\end{equation}
or, if $\sigma(Y_j)$ represents the measurement error in $Y_j$, in a more usual $\chi^2$ form
\begin{equation}
	\Delta_j = \frac{y_j - Y_j}{\sigma(Y_j)}.
\label{eq:witherror}
\end{equation}

If multiple classes ($K$) of constraining observables are used, it is computationally convenient to replace $F$ by
\begin{equation}
	F = \mu S - \frac{1}{2} \sum _k^K \chi ^2 _k.
\end{equation}
The number of observations within each class may be different and may be subject to a different binning schemes.

Our derivation of the weight evolution equation is very similar to that above and is based on constructing a likelihood function giving the likelihood of the model reproducing the measured galactic observations and then maximising it (in log form) subject to a time derivative constraint on the relative entropy of the weights.

Redefining $F$ now as
\begin{equation}
	F = - \frac{1}{2}\chi ^2 + \frac{1}{\epsilon}\frac{dS}{dt} + \mu S,
\label{eqn:simplef}
\end{equation}
and substituting for $S$ gives
\begin{eqnarray}
	\lefteqn{F = - \frac{1}{2}\chi ^2 -  \frac{1}{\epsilon}\frac{d}{dt} \left [ \sum _i^N w_i(t) \ln \left(\frac{w_i(t)}{m_i} \right) \right]}\\
	& & \mbox{} - \mu \sum _i^N w_i(t) \ln \left( \frac{w_i(t)}{m_i} \right).\nonumber
\label{eqn:newf}
\end{eqnarray}
Maximising $F$ with respect to the weights gives the \citet{Syer1996} weight evolution equation
\begin{eqnarray}
	\lefteqn{\frac{d}{dt}w_i(t)=-\epsilon w_i(t) \Biggl [ \sum _j^J \frac{K_j(\bmath{r}_i(t),\bmath{v}_i(t))}{\sigma(Y_j)}\Delta_j(t)}\\
	& & \mbox{} + \mu \left(\ln (\frac{w_i(t)}{m_i}) + 1 \right) \Biggr ].\nonumber
	\label{eqn:rjlweight}
\end{eqnarray}

The M2M method is similar to a maximum entropy method (for example, \citealt{Richstone1988}) but in this case the prior reflects the need for the particle weights to be constant over time.

Additional observational or modelling constraints may be included simply by modifying $F$.  Whether the weights will converge or whether the observations will be reproduced requires either experimentation or a convergence analysis to be performed as per \citet{Syer1996}.  In general, constraints which are expressed as squared `distance measures' appear to perform satisfactorily.  A simple extension to the method which meets these caveats is to amend the $\chi^2$ term to take the form
\begin{equation}
	\frac{1}{2}\chi ^2 _{\rmn{LM}} = \frac{1}{2} \sum _k ^K \lambda _k \chi_k ^2,
\label{eqn:chifactor}
\end{equation}
where the $\lambda_k$ are small positive parameters and may be used to rescale $\chi_k ^2$, or to express the relative priority of observable class $k$ within the M2M model.  Note that, while not expressing $\chi ^2$ as in equation \ref{eqn:chifactor}, \citet{DL2007} employ a scaling factor, not dissimilar in function to the $\lambda _k$, in their `force of change' equation (their $\epsilon ''$).  

Similarly, given that a weight is a fraction of the total galaxy luminosity $L$, that is
\begin{equation}
	\sum _i^N L w_i = L,
\end{equation}
then it is not unreasonable to require that the method does not alter the total luminosity and an appropriate constraint to include is the following (minimisation) term in $F$
\begin{equation}
	- \frac{1}{2} \lambda _{\rmn{sum}} \left( \sum _i^N w_i - 1 \right) ^2.
\label{eqn:sumwt}
\end{equation}
Re-normalisation of the weights was considered and rejected as it would destroy the smoothing history built up in the $\tilde{\Delta}_j$.  \citet{Syer1996} and \citet{DL2007} contain no such similar constraint.  \citet{Dehnen2009} has an alternative scheme for modifying $F$ to achieve weight conservation.

Discrete observables (for example, measurements of the line-of-sight velocities of individual stars) are incorporated into the method as follows.  The probability, $p_{D,j} = p_{D,j}(\bmath{x}_{\perp j}, v_{\parallel j})$, of the model reproducing a discrete line-of-sight velocity measurement is found by convolving the line-of-sight velocity distribution (losvd) with a Gaussian incorporating the observational errors $\sigma _j$.
\begin{equation}
	p_{D,j} = \frac{1}{\sqrt{2\pi} \sigma_j} \int \rmn{losvd}(\bmath{x}_{\perp j}, v_{\parallel}) \exp\left(-\frac{\left(v_{\parallel}-v_{\parallel j}\right)^2}{2\sigma _j^2} \right) dv_{\parallel},
\end{equation}
where the $\parallel$ and $\perp$ subscripts indicate parallel and perpendicular to the line of sight.  Equation  \ref{eqn:newf} gains an additional constraint term
\begin{equation}
	L_{D} = \lambda _D \sum _j ^D \ln (p_{D,j}),
\end{equation}
where $\lambda _D$ is a small positive parameter.  Note that this is not the only way of incorporating discrete observables.  For example, the $p_{D,j}$ could have been included directly in the log likelihood function used to create $\chi ^2$.

From expressing the line-of-sight velocity distribution in terms of the distribution function
\begin{equation}
	\rmn{losvd}(\bmath{x}_{\perp j}, v_{\parallel}) = \frac{\int dx_{\parallel} d^2 \bmath{v}_{\perp} f(\bmath{x}, \bmath{v})}{\int dx_{\parallel} d^3 \bmath{v} f(\bmath{x}, \bmath{v})},
\end{equation}
$p_{D,j}$ is calculated from the model as
\begin{equation}
	p_{D,j} = \frac{1}{\sqrt{2 \pi} \sigma _j} \frac{\sum _i ^N \delta_{ij} w_i \exp\left(-\frac{\left(v_{\parallel i}-v_{\parallel j}\right)^2}{2\sigma _j^2} \right) }{\sum _i ^N \delta_{ij} w_i},
	\label{eqn:indivobs}
\end{equation}
where the selection function $\delta_{ij}$ takes the value $1$ if particle $i$ contributes to observation $j$
and is $0$ otherwise.  The contribution to the weight evolution equation is found by differentiating the log likelihood function $L_D$ with respect to the particle weights and independently exponentially smoothing the resulting numerator and denominator. \citet{DL2008a} arrive at an equivalent expression.  Proper motion data could be incorporated into the method in a similar fashion but this is not explored further here.

If we require a model with an isotropic velocity dispersion, this may be achieved by defining a model observable
\begin{equation}
	y_j = \frac{\sum _i ^N L w_i \left ( 2 v _{r,i} ^2 - v _{t,i} ^2 \right ) \delta _{ij}}{\sum _i ^N L w_i \delta _{ij}},
\label{eqn:iso1}
\end{equation}
where $v_r$ and $v_t$ are the radial and tangential velocities, and including
\begin{equation}
	- \frac{1}{2} \lambda _{\rmn{iso}} \sum _j y _j ^2
\label{eqn:iso2}
\end{equation}
as a minimisation term in $F$.  These expressions come directly from using luminosity weighted velocity dispersions in calculating the $\beta$ anisotropy parameter
\begin{equation}
	\beta = 1 - \frac{\overline{v_t^2}}{2 \overline{v_r^2}}
\end{equation}
and then setting $\beta = 0$.   The denominator $\sum _i ^N L w_i \delta _{ij}$ either should be exponentially smoothed as per equation \ref{eq:tempsmooth} to reduce particle counting effects, or alternatively may be replaced by an observationally derived value by recognising that $L_j = \sum _i ^N L w_i \delta _{ij}$ is the luminosity of radial shell $j$. There is a third option which is to create a composite constraint of luminosity times the original constraint.  The $L_j$ approach is used for the isotropy constraint in the rest of this paper.  These 3 options apply not just to the isotropy constraint but to all constraints where a sum of weights appears in the denominator. To be precise, the isotropy constraint is a constraint on $\beta$ and it will not enforce strict isotropy with the same dispersion in each of the 3 velocity components $v_r$, $v_{\theta}$ and $v_{\phi}$.  Extending the constraint to the case where $\beta = \beta (r)$ is straightforward and has been implemented in \citet{Dehnen2009}. 

\subsection{Model observables and kernels}\label{sec:obskernels}
In this section we list the model observables and kernels from which we use an appropriate subset in the M2M models described in this paper.  We show the selection function separately from the actual kernel and also abbreviate $K_j(\bmath{r}_i (t), \bmath{v}_i (t))$ to $K_{ji}$.  Within a given class of observables, one particle contributes to only one observable.  No attempts have been made to smear particles to mimic the effects of an observational point-spread function, say, as in \citet{Syer1996} and \citet{DL2007}. The total luminosity of the galaxy being modelled is $L$.  

\begin{enumerate}
\item Luminosity density 
\begin{eqnarray}
y_j & = & \sum _i ^N \delta _{ij} \frac{L w_i}{V_j}\\
K_{ji} & = & \frac{L}{V_j}
\end{eqnarray}
where $V_j$ is the volume of the associated model bin.

\item Surface brightness
\begin{eqnarray}
y_j & = & \sum _i ^N \delta _{ij} \frac{L w_i}{A_j}\\
K_{ji} & = & \frac{L}{A_j}
\end{eqnarray}
where $A_j$ is the area of the associated model bin.

\item \label{item:sblos}Surface brightness times luminosity-weighted line-of-sight second velocity moment 
\begin{eqnarray}
y_j & = & \sum _i ^N \delta _{ij} L w_i \frac{v^2_{\parallel i}}{A_j}\\
K_{ji} & = & \frac{L v^2_{\parallel i}}{A_j}
\end{eqnarray}

\item Luminosity-weighted line-of-sight second velocity moment
\begin{eqnarray}
y_j & = & \sum _i ^N \delta _{ij} L w_i \frac{v^2_{\parallel i}}{A_j I_j}\\
K_{ji} & = & \frac{L v^2_{\parallel i}}{A_j I_j}
\end{eqnarray}
where $I_j$ is the measured surface brightness.  This form of the observable is an alternative to that in item \ref{item:sblos}.

\item Surface luminosity times line-of-sight velocity distribution Gauss-Hermite coefficient ($n$)
\begin{eqnarray}
y_j & = & \sqrt 2 \gamma ^{-1} L \sum _i ^N \delta _{ij} w_i H_n(v_{\rmn{norm},i}) \exp(- v_{\rmn{norm},i}^2 / 2)\\
K_{ji} & = & \sqrt 2 \gamma ^{-1} L H_n(v_{\rmn{norm},i}) \exp(- v_{\rmn{norm},i}^2 / 2)
\end{eqnarray}
where $v_{\rmn{norm},i} = (v_{\parallel i} - v_{\rmn{best}}) / \sigma _{\rmn{best}}$. $\gamma$, $v_{\rmn{best}}$ and $\sigma _{\rmn{best}}$ are the line strength, the mean line-of-sight velocity and dispersion from the best Gaussian fit to the observed line-of-sight velocity distribution data.  To determine the Hermite polynomial values, we use the recurrence relationship
\begin{equation}
H_p(x) = \sqrt{\frac{2}{p}} x H_{p-1}(x) - \sqrt{\frac{p-1}{p}}H_{p-2}(x), \; \; \; \; \; \; \; p \geq 2
\label{eqn:hermiterecur}
\end{equation}
with $H_0(x) = 0$ and $H_1(x) = \sqrt{2} x$.

Modelling the line-of-sight velocity distribution with a truncated Gauss-Hermite polynomial series is discussed in \citet{Gerhard1993} and \citet{Vandermarel1993}.
\end{enumerate}

We have not at this time implemented any schemes for deprojecting surface brightness to give a luminosity density distribution as in \citet{EJ2007} or \citet{DL2008a} where a multi-Gaussian expansion (\citealt{Emsellem1994}) is used to construct the distribution for their elliptical galaxies. Deprojection is not an integral part of M2M method.  If we require a luminosity density, we assume that deprojection could be performed and just use a theoretical luminosity density, modified as in section \ref{sec:finitemodel}.

\section{Implementation}\label{sec:implementation}
We describe in this section some of the key practical issues to be addressed in designing and implementing a software system to meet the theoretical design in section \ref{sec:m2m}.
 
\subsection{Process and data flows}
We have decomposed the process and data flows into 3 main phases.  The first phase, preparation, covers creating the particle initial conditions, and either creating from theoretical functions the observational constraints to be used, or manipulating observations of real galaxies into a form suitable for modelling.  The second phase is the actual running of the M2M model, and the third phase is concerned with the analysis of the output from the modelling run.  The analysis phase is split into 3 components namely particle weight convergence, reproduction of the observational constraints and particle kinematics.  

\begin{figure}
		\includegraphics[width=84mm]{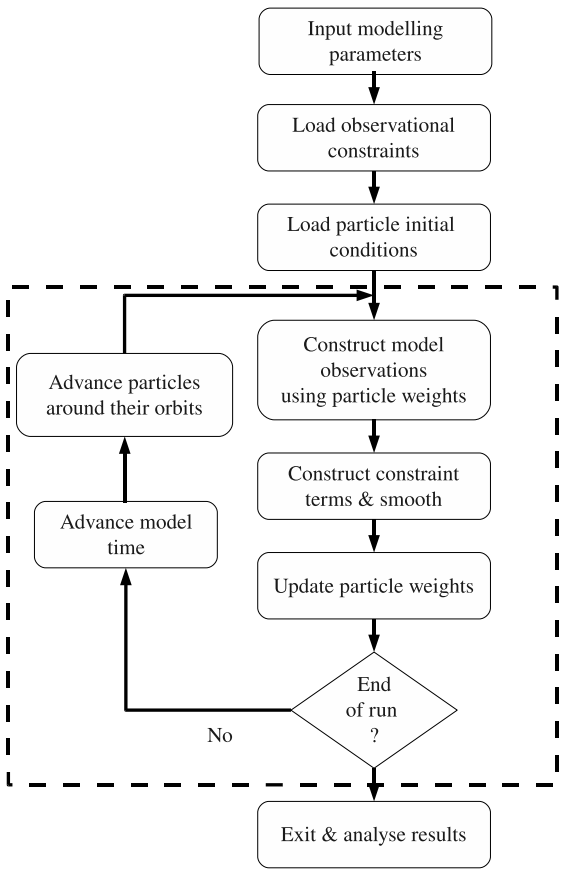}
		\caption{Execution phase flowchart.  The parallelised processes are contained within the dashed box.}
	\label{fig:flowchart}
\end{figure}

The execution phase follows quite naturally from the equations in section \ref{sec:m2m} and is shown in Figure \ref{fig:flowchart}.  Given the number of particles used in modelling runs, in our case $5 \times 10^4 - 2 \times 10^6$, we have parallelised our implementation so that multiple computer processors may be used to reduce the overall execution elapsed times.  We adopt a simple strategy whereby 1 processor controls the modelling run and is responsible for the collation and smoothing of the model produced observations and calculating the weight evolution constraint terms.  The other processors are responsible for orbiting their subset of the particles, calculating their contribution to the model observations and updating their particle weights.  At the end of a run, the control processor collates all the particle data for subsequent analysis.  We find that this parallelisation strategy, without change, works acceptably well across a range of hardware platforms from single dual processor machines, clusters of PCs and on high performance computing systems with fast inter-processor data connections.  As shown in Table \ref{tab:scaling}, doubling the number of processors approximately halves the computer elapsed time.  Note that our implementation does not as yet handle self-consistent potentials created by the particles themselves.
\begin{table}
	\begin{center}
	\caption{MPI scaling}
	\label{tab:scaling}
	\begin{tabular}{|c|c|}
			\hline
			\textbf{Processors} & \textbf{Elapsed time (m)} \\
			\hline
			8 & 18.3\\
			16 & 8.0\\
			32 & 4.0\\
			64 & 2.0\\
			\hline
	\end{tabular}
	\end{center}
\medskip
The elapsed time figures are from a $5 \times 10^5$ particle model with 3 observational constraints run for 100 dynamical time units  in a high performance computing environment.
\end{table}

The weight convergence algorithm described in section \ref{sec:wtobs} we have implemented in computer memory with the slave processors being responsible for monitoring the convergence of their own particles.  At the end of a modelling run, the controlling processor collates the weight convergence data for subsequent analysis.  Clearly, alternative schemes are possible, for example using filestore.

\subsection{Finite modelling}\label{sec:finitemodel}
All the models we use in this paper are either spherical or spheroidal.  We set a maximum radius for a model and arrange that particles do not `escape' from the model by setting an energy constraint on their initial velocities ($v$).  That is
\begin{equation}
	v^2 \leq 2 \left[ \phi(\rmn{boundary}) - \phi(\rmn{initial \; position}) \right].
\end{equation}

In modelling galaxies, a system of large spatial extent, possibly infinite if theoretical functions are used, is being modelled in a  computer system.  Inevitably, model boundary effects occur primarily as a result of model observable values falling off faster with radius than would be the case with theoretical functions, and the limited number of particles close to the boundary.  We solve this issue in one of two ways, either by the simple expedient of oversizing the models and only analysing the central portions, or by truncating the distribution function and constructing a tailored observable function.  The distribution function is truncated by energy, limiting the energy to the range between the potential energy at the origin and at the model boundary, and setting the distribution function to zero for energies outside of this range.  As \citet{Kashlinsky1988} points out, this is not the only way of truncating the distribution function.  For example, truncation by radius, resulting in an increase in the number of (bound) circular orbits, may be appropriate but we have conducted no experiments using this approach.

Assuming a distribution function which is a function of energy only, we construct the luminosity density function via a particle realisation of the truncated distribution function over the radial extent of the M2M model.  To ensure the correct radial distribution of particles, we determine the density of particles with a given energy, $\nu _{\rmn{E}} (r)$, by integrating the distribution function over velocity space to give
\begin{equation}
	\nu _{\rmn{E}} (r) = 4 \pi f(E) \sqrt{2 \left( E - \phi (r) \right)}.
\label{eqn:rhoe}
\end{equation}
We obtain the radial position for a particle by sampling uniformly randomly from the `fraction within radius' function, $N _{\rmn{E}} (<r)$ given by
\begin{equation}
	N _{\rmn{E}} (<r) = \frac{16 \pi ^2 \int _0 ^r \sqrt{2 \left( E - \phi (r) \right)} r^2 dr}{g(E)},
	\label{eqn:fractione}
\end{equation}
where $g(E)$ is the density of states function. 

\begin{figure}
	\centering
	\includegraphics[width=84mm]{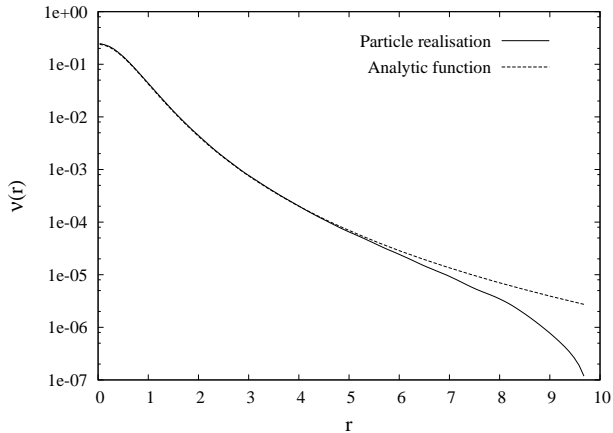}
	\caption[Luminosity density function comparison]{Comparison for a Plummer sphere of the theoretical luminosity density function and the equivalent function from a particle realisation.}
	\label{fig:densityplot}
\end{figure}

Figure \ref{fig:densityplot} compares a luminosity density function created in this way with the usual analytic density function for a Plummer sphere (radii are in units of the core radius).  As can be seen, the two functions match in the inner part of the model but with the constructed function going to zero as required at the boundary of the model.  For data creation purposes, the constructed luminosity density function is used in tabular form and integrated numerically to create other luminosity related functions, for example surface brightness or luminosity weighted velocity dispersion.

\begin{figure}
	\centering
	\includegraphics[width=84mm]{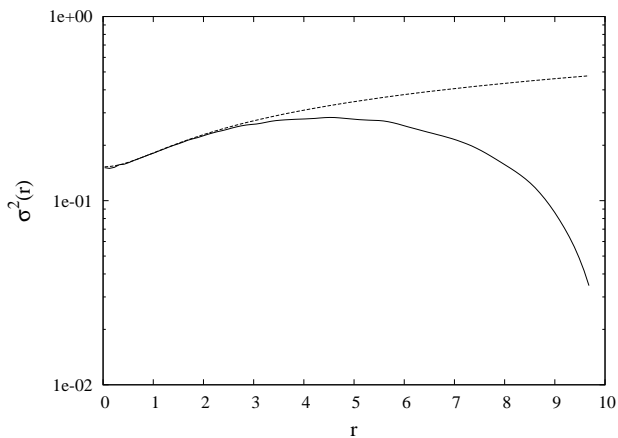}
	\caption[Rising velocity dispersion]{Comparison for a \citet{Wilkinson2002} model, with the potential power law parameter $\alpha = -0.5 $, of the rising theoretical velocity dispersion and the equivalent from a particle realisation.  Only the inner $\approx 25\%$ of the model is usable.}
	\label{fig:risingdispplot}
\end{figure}

Luminosity density is an example of an observable which decreases with radius.  Modelling observables which increase with radius, for example a rising velocity dispersion, the effects are more extreme.  As can be seen from Figure \ref{fig:risingdispplot}, the inner part of the model where there is a good match between the theoretical function and the particle realisation is reduced to $\approx 25\%$ of the radial extent of the model.  

\subsection{Particle initial conditions} \label{sec:particleics}
All particles are given the same initial weight and prior equal to $1/N$, where $N$ is the number of particles in the model.

For the particles' initial spatial and velocity coordinates we use one of three schemes,
\begin{enumerate}
	\item spatial coordinates allocated such that the particle distribution matches the luminosity distribution, and the velocity coordinates uniformly randomly distributed,
	\item spatial coordinates allocated such that the particle distribution matches the luminosity distribution, and the velocity coordinates sampled randomly from a Gaussian distribution created from the velocity dispersion function (from solving the Jeans equations), 
	\item by assuming that the distribution function is a function of relative energy only, spatial and velocity cordinates obtained from from distributing the particles uniformly randomly in energy.
\end{enumerate}
In the third scheme, we sample uniformly, randomly from the integrated differential energy distribution (see \citealt{BT2008}) to obtain a particle's energy, use the energy to determine the maximum radius it implies, allocate the particle's spatial position within that radius using $\nu _{\rmn{E}} (r)$ (equation \ref{eqn:rhoe}), and finally use the energy difference between between the particle's energy and its spatial position to allocate the velocity components.  

The first two schemes are appropriate when observational data from a real galaxy is being used, and the third scheme, when theoretical models are being used.  All the schemes clearly need at least a potential to be specified in order to be used.  We have however deliberately separated the creation of the initial conditions from the main modelling software to increase flexibility - we may choose to run the model with a different potential for example.

\subsection{Observational constraints}\label{sec:gobs}
Clearly any observations of a real galaxy will need to go through a process of manipulation and conversion in order to get them into a form where they can be used with a M2M model.   For theoretical galaxy models, for the purposes of developing the M2M method, we create constraint data by sampling from a Gaussian distribution with mean the function value, and standard deviation calculated from a pre-specified relative error.  We constrain data values so produced by limiting them to be within $n$-sigma of the function values ($n$ is typically $2$ or $3$).

\subsection{Weight convergence \& observable reproduction}\label{sec:wtobs}
In assessing whether or not a modelling run has been successful, we require, amongst other criteria, a high degree of weight convergence and  consistent observable reproduction over a number of orbits.  In practice, we choose some representative time period $T_c$ which will cover, for example, both the long and short period particle orbits.  For weight convergence, we consider a particle's weight to have converged if its maximum relative deviation from its mean weight over time period $T_c$ is less than some predetermined model-wide tolerance, that is if
\begin{equation}
	\rmn{max} \left \vert \frac{w_i - w_{\rmn{mean},i}}{w_{\rmn{mean},i}} \right \vert \leq \rmn{tolerance}.
\end{equation}
We take as our unit of time the local dynamical time at the half mass radius of the model
\begin{equation}
	1\: \rmn{time \: unit} = \sqrt{\frac{3 \pi}{16 G \tilde{\rho}}},
\label{eqn:timeunit}
\end{equation}
where $\tilde{\rho}$ is the mean density inside the half mass radius.  For a typical model run, we set $T_c$ to be $\approx 10\%$ of the model duration and use a tolerance of $5$ per cent or less.

Good model stability with respect to the particle weights is also required.  Stability in this context means the degree to which a model's outputs vary if the particles are subsequently orbited with weight evolution turned off.  This is particularly important if the particles are to be used in some further modelling process. Stability is affected by the number of particles with converged weights and the total weight associated with particles with unconverged weights, and is illustrated in section \ref{sec:basicmodelcap}.

We consider a M2M model to have reproduced the actual observations if the smoothed model observations match the actual observations to within the measurement errors on the actual observations, that is if
\begin{equation}
	\left \vert \tilde{y} _j - Y_j \right \vert \leq \sigma (Y_j).
\end{equation}
$\tilde{y} _j$ may be calculated directly from $\tilde{\Delta} _j$ using the smoothed form of equation \ref{eq:witherror}.  Note that observable non-reproduction does not necessarily imply that the method has failed - the smoothness of the constraining data needs to be taken into account as does the degree of smoothing employed in the model.

Given our success criteria for a model run and the mechanisms we use to measure whether they have been achieved, we do not deploy a `phase mixing' phase as described in \citet{DL2007}.

\subsection{Numerical methods}
For orbit integration, we use the standard interleaved second order leapfrog method (drift, kick, drift) with either a fixed or adaptive time step. We either specify the time step  directly or allow the model to determine it from the local dynamical time at the origin.  For adaptive time stepping we use a 3 level model, progressively decreasing a particle's time step by a factor of $2$ as it nears the origin. Over the duration of a typical modelling run of $250$ dynamical time units, we achieve an average maximum relative energy precision $\Delta E(t) / E(0) \approx 10^{-4}$  which represents a reasonable level of energy conservation and is satisfactory for M2M purposes.  We have tried (and discarded pending further investigation) the `dimensionless time' orbit integration noted in \citet{Dehnen2009}.   While it does increase the numbers of orbits of particles with long orbital periods, we find that weight convergence is reduced by $\approx 2\%$ and that there is no change regarding the type of particles with unconverged weights at the end of a modelling run - it remains the highest energy particles.

For integration of the weight evolution equation and exponential smoothing we use Euler's method.  For example, the exponential smoothing equation (equation \ref{eq:tempsmooth}) becomes
\begin{equation}
\label{eqn:numsmooth}
	\tilde{\Delta}_j(t+1) = \tilde{\Delta}_j(t) + \alpha \delta t \left[\Delta _j(t) - \tilde{\Delta}_j(t) \right].
\end{equation}
Rather than updating the particle weights every time step, we take the weight evolution time step as an integer multiple ($\leq 5$ for production runs) of the orbit integration time step.  By reducing the inter-processor data traffic, a saving in computer elapsed time is achieved with no loss in effectiveness of the M2M method.  For example, changing the multiple from $1$ to $5$ results in $\approx 50\%$ saving in elapsed time with minimal changes in both weight convergence and the model $\chi ^2 _{\rmn{LM}}$ value. 

For three dimensional radially dependant observables, we use a radial binning scheme, and a polar scheme, with azimuthal binning if required, for two dimensional observables. The bin sizes may be either regular or irregular. Also, individual bins are only used if required by the distribution of the observational data - that is, we allow for gaps in the data. For the regular schemes, we use uniform or logarithmic bin sizes, or, to assist in the resolution of any central density peak, pseudo-logarithmic radial bin sizes as described in \citet{Sellwood2003}.  
\begin{equation}
	r_i = (r_{\rmn{max}} + 1)^{i/B} - 1,\; \; \; \; \; i=1, \cdot \cdot \cdot ,B
\end{equation}
where $r_{\rmn{max}}$ is the maximum radius, divided into $B$ bins, and $r_i$ is the radius of the $i^{th}$ bin boundary.  The software for binning, and also that for numerical integration and interpolation, utilises the GNU Scientific Library\footnote{http://www.gnu.org/software/gsl/}.

\section{Creating a M2M Model}\label{sec:basicmodelcap}
In this section, we consider various aspects of creating a M2M model and cover tuning its key parameters, the impact of different particle initial conditions, how many particles to use, and modelling incomplete data sets.  Section \ref{sec:entropy} deals specifically with the impact of the relative entropy derivative constraint.

\subsection{Parameter tuning}\label{sec:paramtuning}
The parameters are the weight convergence rate $\epsilon$, $\alpha$ for exponential smoothing, $\mu$ governing regularisation, the observable constraint $\lambda _k$'s from equation \ref{eqn:chifactor}, and $\lambda _{\rmn{sum}}$ and $\lambda _{\rmn{iso}}$ for the sum of weights constraint and isotropic dispersion constraint.  The parameters are treated as tunable with their values being determined prior to the start of a modelling run.  Setting the parameters should be thought of as a process - we have identified no simple mechanism which will determine all the parameters required in one trial modelling run.  The key to the process is to establish the initial values for the parameters.  Having done this, it is straightforward to perform a series of modelling runs, increasing or decreasing the parameter values, to achieve a particular desired position.  

We first explain how we assess the impact of the relative entropy (regularisation) term in the weight evolution equation.  To recap, where constructed observable data are being used, we create the data using Gaussian sampling from the various functions of the associated mathematical model. The combination of relative error and sigma cut-off determine the spread of data points around the functions.  In practice, what is needed from a M2M model, no matter whether the data is constructed or real, is not that the observable data points are individually reproduced but that a smooth curve approximating the data points and reflecting any key features in the data is generated. Given the method of data construction, the smooth curve approximating the data points should be the function the data were generated from.  To give a measure of how effective a M2M model is in re-creating the underlying functions, the following relative sum of squares is calculated for every class $s$ of observable constraint.
\begin{equation}
	C_s = \sum ^J _j \left ( \frac{x_j - X_j}{X_j} \right ) ^2
\label{eqn:smoothsq}
\end{equation}
where $X_j$ is the theoretical value for data point $j$ and $x_j$ is the equivalent smoothed value produced by the M2M model
\begin{equation}
	x_j = Y_j + \tilde{\Delta}_j \sigma(Y_j).
\end{equation}
Where observations of a real galaxy are being used, the underlying functions will not be known.  The solution then is to create a trial data set approximating the real data set, with known underlying functions, and to use that trial data to determine the degree of regularisation required (the value of $\mu$) when the real data is modelled.  Note also that regularisation is model-wide and is not specific to any one constraint.  Our preference is to start with a low value of $\mu$ initially ($\mu \approx 10^{-3}$).  We find that as $\mu$ is increased it becomes necessary to reduce $\epsilon$ to maintain good weight convergence.

Given the above, our completion criteria for a tuning exercise are that a high degree of weight convergence has been achieved, that the amount of unconverged weight is low, that the observable $C_s$ values described above are stable (orbiting the particles with weight evolution turned off does not cause significant change), and that all other constraints have been met.

For the observable constraints, the weight evolution equation contains terms of the form
\begin{equation}
	\frac{K_{ji} \tilde{\Delta} _j}{\sigma (Y_j)}.
\end{equation}
In the following, we quantify the impact of the term numerically and relate it to the value of the relevant $\lambda _k$ parameter and to the relative strengths of the observable constraint terms in the weight evolution equation.

Replacing  $\tilde{\Delta} _j$ by its unsmoothed form and replacing $\sigma (Y_j)$ by
\begin{equation}
	\sigma (Y_j) = \alpha _j Y_j
\end{equation}
(where $\alpha _j$ is the fractional error) gives terms of the form
\begin{equation}
	\frac{K_{ji}}{\alpha _j ^2 Y_j} \frac{y_j - Y_j}{Y_j}.
\end{equation}
Assuming the right hand fraction is numerically comparable for all observables then it is the left hand fraction which dictates the constraint contribution to the weight evolution equation.  

For example, using the expressions for the kernels in section \ref{sec:obskernels} and ignoring the total luminosity of the galaxy being modelled, the left hand fractions (evolution factors) for the surface brightness and dispersion constraints are
\begin{eqnarray}
	T_{\rmn{SB},j}  & = & \frac{1}{\alpha _j ^2 Y_j A_j} \\
	T_{\rmn{VD},j}  & = & \frac{v^2_{\parallel _j}}{\alpha _j ^2 Y_j A_j}
\end{eqnarray}
where $v_{\parallel j}$ is some line of sight velocity.  For simplicity, $v_{\parallel j}$ is taken as the maximum velocity occurring in the model.  When comparing model runs at different mass-to-light ratios, given that $v^2_{\parallel _j}$ scales according to the ratio $\Upsilon$, the balance of the terms in the weight evolution equation will change across the runs.  This can be resolved by multiplying $T_{\rmn{VD},j}$ by $\Upsilon ^{-1}$.

From experience, we have found suitable start values for the observable constraint parameters by making the product of the typical evolution factor and parameter $\approx O(1)$, that is, the parameter is used to neutralise the evolution factor.  For example, for surface brightness
\begin{equation}
	T_{\rmn{SB},j} \lambda _{\rmn{SB}} \approx O(1)
\end{equation}
Clearly, the evolution factors will have a range of values perhaps spanning several orders of magnitude.  For the models in this paper, taking the value at $2$ effective radii gives a suitable compromise.

Using a similar analysis to that above for the sum of weights and isotropic dispersion constraints, we set the product of the typical evolution factor and parameter to be $\approx 10^{-2}$.  Setting the values initially to be lower than for the observable constraints means that observable constraints have a stronger influence in the weight evolution equation.

For the exponential smoothing parameter $\alpha$, we perform a series of modelling runs with no regularisation and with $\alpha$ being varied from $10^1$ (no smoothing, $\alpha = \delta t ^{-1}$ - see equation \ref{eqn:numsmooth}) to $5 \times 10 ^{-3}$.  We find that, as the amount of exponential smoothing increases, weight convergence increases, the unconverged weight decreases and the magnitude of the $\chi ^2 _{\rmn{LM}}$ gradient increases.  The statistical fluctuations in $\chi ^2 _{\rmn{LM}}$ are greatly reduced once $10^{-2} < \alpha < 10^{-1}$ where the exact value depends on the number of particles being used ($5 \times 10^4$ at the lower end to $10^6$ at the upper).  It therefore turns out that the \citet{Syer1996} value of $\alpha = 5.24 \times 10^{-2}$ is in fact a reasonable default value to use over quite a wide range in particle numbers and bin configurations.

To summarise,
\begin{enumerate}
\item Parameter determination must be viewed as a process.
\item For the main observable constraints, setting the product of the typical evolution factor and parameter to be of $O(1)$ gives a means of determining the initial values of the parameters.  For other $\lambda$ constraints, setting the product to $\approx 10^{-2}$ gives a usable start position.
\item For the other parameters, $\epsilon = 0.025$, $\alpha = 0.05$ and $\mu = 0.001$ are reasonable starting values.
\item The larger the observable errors and the greater the spread of data points, the more likely it is that smoothing / regularisation from the relative entropy term will be needed and a higher value of $\mu$ required.  Quite how much regularisation is required (or desirable) is application specific.  With increased regularisation it is highly likely that observable reproduction will reduce. Based on the experiments we have conducted, a higher value of $\mu$, changing the influence of terms in the weight evolution equation, requires a reduction in the value of $\epsilon$ to achieve acceptable results (for example, high weight convergence). 
\item Regardless of the value of $\epsilon$, the rate of particle weight convergence should be monitored.  Usable results may be obtainable from shorter modelling runs.
\end{enumerate}

As a final comment, particularly when comparing parameter values between papers, it is the $\epsilon$ parameter product (for example, $\epsilon \mu$ in equation \ref{eqn:newf}) which is important.  Parameter values may be rescaled as required provided $\epsilon$ is rescaled correspondingly.  Also, determination of the parameter values may appear onerous but it should be remembered that it is only necessary to include the constraints and parameters that are required.   The \citet{Syer1996} $\mu$ adjustment process has not been used but the process or its equivalent may be applicable to other parameters and requires further investigation.

\subsection{Relative entropy and the derivative constraint}\label{sec:entropy}
We investigate the behaviour of the relative entropy regularisation term ($S$) and the relative entropy derivative constraint ($dS/dt$) as the $\mu$ parameter is increased from $10^{-3}$ to $10^3$.  We perform 2 sets of runs, the first having no further constraints and the second using the total particle weight constraint.  $5 \times 10^4$ particles are used and $\lambda_{\rmn{sum}} = 7 \times 10^2$ in the second set of runs.  
\begin{table}
	\begin{center}
	\caption{Regularisation}
	\label{tab:rawentropy}
	\begin{tabular}{ccccc}
	\hline
	& \multicolumn{2}{c}{\textbf{Unconstrained Weight}} & \multicolumn{2}{c}{\textbf{Constrained Weight}}\\
	$\bmath{\mu}$ & $\bmath{S}$ & $\bmath{\sum w_i}$ &  $\bmath{S}$ & $\bmath{\sum w_i}$ \\
	\hline
	$1.0 \: 10^{-3}$ & $3.52 \: 10^{-3}$ & $1.00$ & $1.43 \: 10^{-6}$ & $1.00$ \\
	$1.0 \: 10^{-2}$ & $3.36 \: 10^{-2}$ & $0.97$ & $1.43 \: 10^{-5}$ & $1.00$ \\
	$1.0 \: 10^{-1}$ & $2.21 \: 10^{-1}$ & $0.74$ & $1.43 \: 10^{-4}$ & $1.00$ \\
	$1.0 \: 10^{0}$  & $3.68 \: 10^{-1}$ & $0.38$ & $1.43 \: 10^{-3}$ & $1.00$ \\
	$1.0 \: 10^{1}$  & $3.68 \: 10^{-1}$ & $0.37$ & $1.40 \: 10^{-2}$ & $0.99$ \\
	$1.0 \: 10^{2}$  & $3.68 \: 10^{-1}$ & $0.37$ & $1.16 \: 10^{-1}$ & $0.88$ \\
	$1.0 \: 10^{3}$  & $3.68 \: 10^{-1}$ & $0.37$ & $3.41 \: 10^{-1}$ & $0.52$ \\
	\hline
	\end{tabular}
	\end{center}
\medskip
Increasing regularisation ($\mu$) with and without the total weight constraint, and no other constraints.
\end{table}
The theoretical maximum in $S$ occurs when $w_i = m_i/e$ at which time, given that $m_i = 1/N$ where $N$ is the number of particles, $S_{\rmn{max}} = 1/e$ and $\sum w_i = 1/e$. For higher value of $\mu$, these values for $S_{\rmn{max}}$ and $\sum w_i$ can be seen in Table \ref{tab:rawentropy} in the `unconstrained weight' columns.  However, there is a conflict between the value of $\sum w_i$ at $S_{\rmn{max}}$ and the requirement for $\sum w_i = 1$ to conserve the luminosity of the galaxy being modelled.  Imposing the total weight constraint causes the requirement to be met but at a lower value of $S_{\rmn{max}}$ (the `constrained weight' columns of Table \ref{tab:rawentropy}).  Where $\sum w_i \neq 1$, this can be resolved by increasing the value of $\lambda_{\rmn{sum}}$.  For both sets of runs, $dS/dt$ does tend to zero with increasing time, with the timescale to do so reducing as $\mu$ is increased.
Within the maximisation of $F$, equation \ref{eqn:simplef}, $dS/dt$ behaves as the constraint $dS/dt = 0$ and not as a function to be maximised.

\begin{figure}
	\includegraphics[width=84mm]{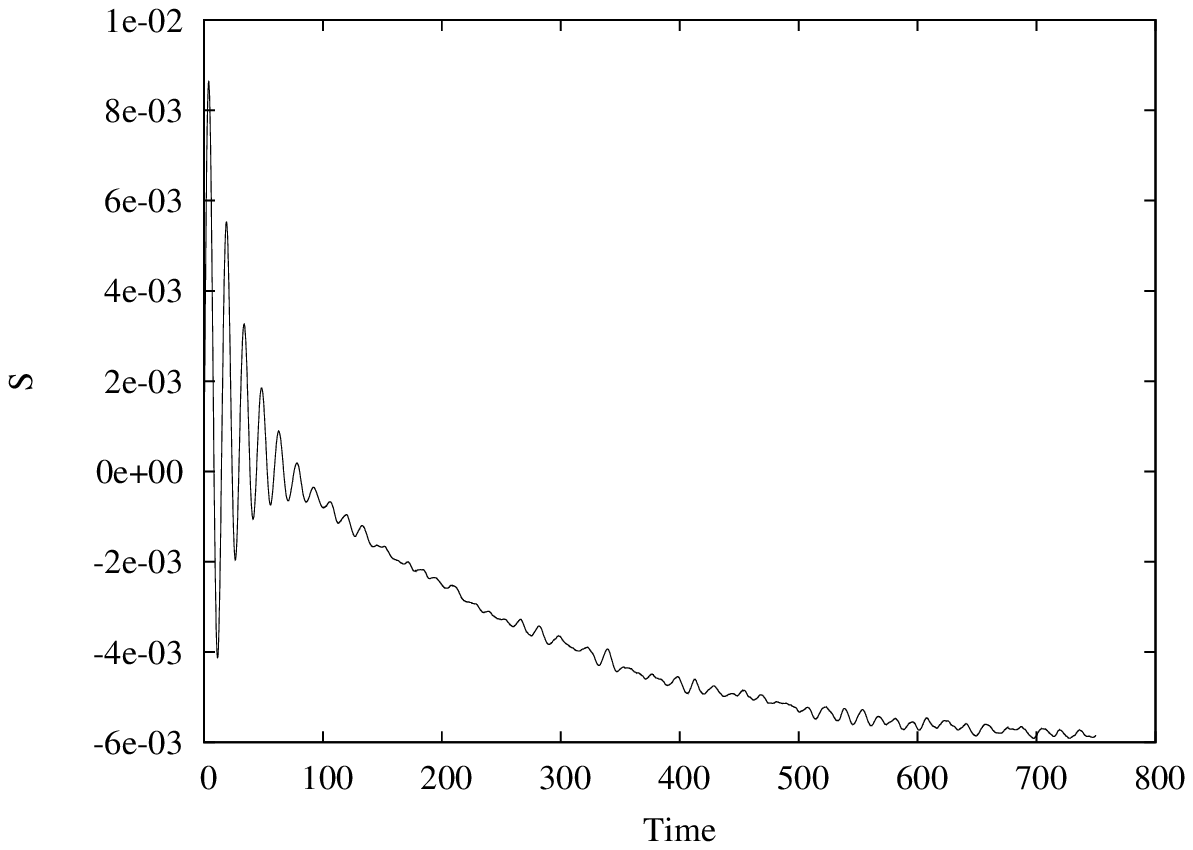}
	\includegraphics[width=84mm]{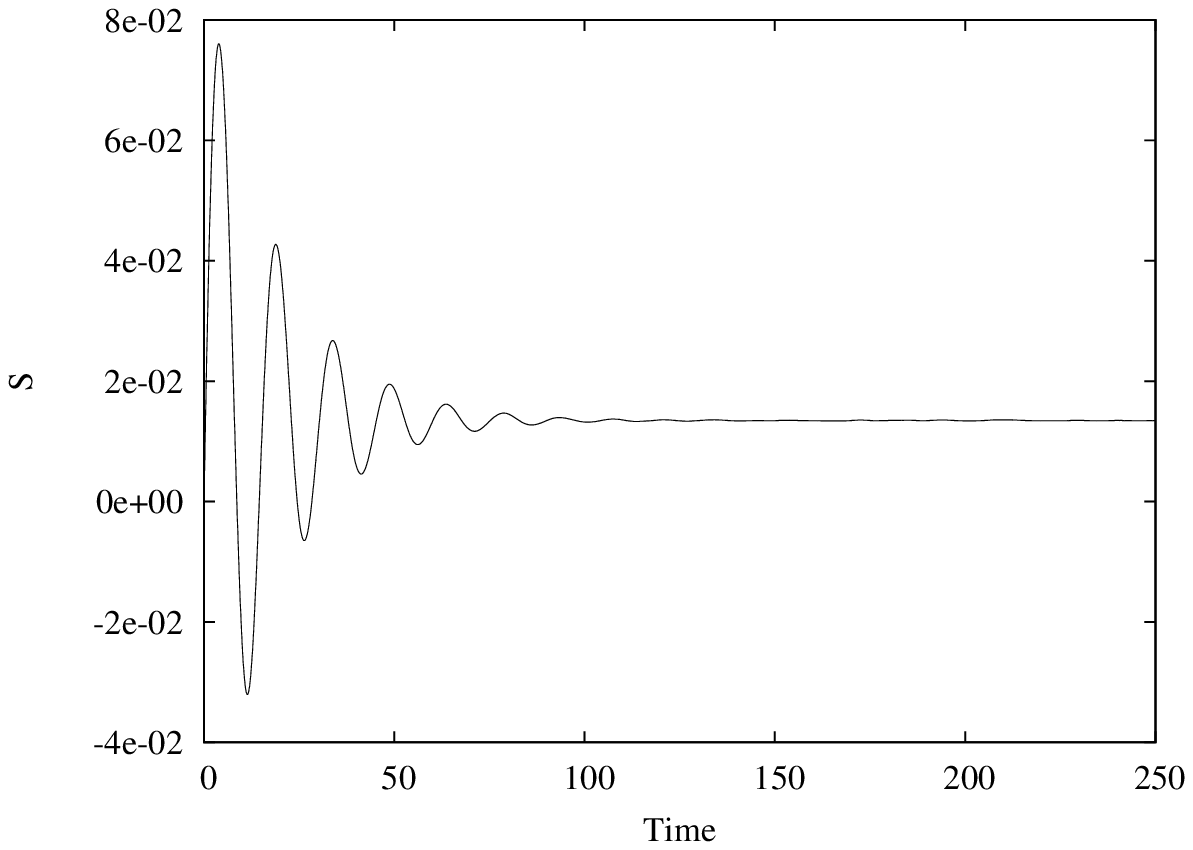}\\
\caption{Relative entropy time evolution for $\mu = 1.0$ (top panel) and  $\mu = 10.0$ (bottom panel).  A relative entropy constant value is achieved after $\approx 750$ time units for the lower value of $\mu$ and after $\approx 100$ units for the higher value.}
\label{fig:compentflat}
\end{figure}

We now extend the model to include other constraints (as in Table \ref{tab:icmodel}) and the results are recorded in Table \ref{tab:entropy}.  For completeness, we also include a run with $\mu = 0$, that is, with no regularisation.
\begin{table*}
	\begin{center}
	\caption{Impact of increasing regularisation}
	\label{tab:entropy}
	\begin{tabular}{ccccccccccc}
		\hline
		$\bmath{\mu}$ & $\bmath{-F}$ & $\bmath{S}$ & $\bmath{\chi ^2 _{\rmn{LM}}}$ & $\bmath{\chi ^ 2 _{\rmn{SB}}}$ & $\bmath{\chi ^ 2 _{\rmn{V2}}}$ & \textbf{Particles} &\textbf{Uconv} & $\bmath{\sum w_i}$ & $\bmath{C _{\rmn{SB}}}$ & $\bmath{C _{\rmn{V2}}}$ \\
	& & & &  &  & \textbf{Converged} & \textbf{Weight} \\
	& & & &  &  & \textbf{(\%)} & \textbf{(\%)} \\
	\hline
	$0.0$            & $5.48 \: 10^{-2}$ & $-9.03 \: 10^{-3}$ & $1.10 \: 10^{-1}$ & $2.27 \: 10^1$ & $1.88 \: 10^1$ & $99.53$ & $0.62$ & $1.00$ & $3.01 \: 10^{-2}$ & $1.63 \: 10^{-1}$\\
	$1.0 \: 10^{-3}$ & $5.48 \: 10^{-2}$ & $-9.02 \: 10^{-3}$ & $1.10 \: 10^{-1}$ & $2.27 \: 10^1$ & $1.88 \: 10^1$ & $99.53$ & $0.62$ & $1.00$ & $3.01 \: 10^{-2}$ & $1.63 \: 10^{-1}$\\
	$1.0 \: 10^{-2}$ & $5.49 \: 10^{-2}$ & $-8.95 \: 10^{-3}$ & $1.10 \: 10^{-1}$ & $2.28 \: 10^1$ & $1.88 \: 10^1$ & $99.53$ & $0.63$ & $1.00$ & $3.00 \: 10^{-2}$ & $1.63 \: 10^{-1}$\\
	$1.0 \: 10^{-1}$ & $5.61 \: 10^{-2}$ & $-8.26 \: 10^{-3}$ & $1.10 \: 10^{-1}$ & $2.30 \: 10^1$ & $1.90 \: 10^1$ & $99.54$ & $0.62$ & $1.00$ & $2.93 \: 10^{-2}$ & $1.57 \: 10^{-1}$\\
	$1.0 \: 10^{0}$  & $6.27 \: 10^{-2}$ & $-3.28 \: 10^{-3}$ & $1.17 \: 10^{-1}$ & $2.51 \: 10^1$ & $2.04 \: 10^1$ & $99.44$ & $0.72$ & $1.00$ & $2.43 \: 10^{-2}$ & $1.13 \: 10^{-1}$\\
	$1.0 \: 10^{1}$  & $5.00 \: 10^{-3}$ & $1.34 \: 10^{-2}$  & $1.40 \: 10^{-1}$ & $3.48 \: 10^1$ & $2.44 \: 10^1$ & $98.23$ & $1.96$ & $0.99$ & $1.23 \: 10^{-2}$ & $2.63 \: 10^{-2}$\\
	$1.0 \: 10^{2}$  & $-6.07 \: 10^{0}$ & $1.51 \: 10^{-1}$  & $2.64 \: 10^{-1}$ & $2.32 \: 10^2$ & $2.93 \: 10^1$ & $96.64$ & $3.53$ & $0.88$ & $3.99 \: 10^{-1}$ & $3.02 \: 10^{-1}$\\
	$1.0 \: 10^{3}$  & $-2.58 \: 10^{2}$ & $3.41 \: 10^{-1}$  & $2.31 \: 10^{0}$  & $3.13 \: 10^3$ & $1.44 \: 10^2$ & $99.03$ & $1.01$ & $0.52$ & $7.29 \: 10^{0}$  & $5.47 \: 10^{0}$\\
	\hline
	\end{tabular}
	\end{center}
\medskip
Increasing regularisation ($\mu$) with the total weight constraint applied and other constraints as in Table \ref{tab:icmodel}.  $C _{\rmn{SB}}$ and $C _{\rmn{V2}}$ are defined as per equation \ref{eqn:smoothsq}.  At higher values of $\mu$, the weight evolution equation becomes `unbalanced'. $S$ increases as do the constraint $\chi ^2$ values.  
\end{table*}
As before, the time by which $dS/dt \approx 0$ and $S \approx \rmn{constant}$ reduces at higher $\mu$ values (illustrated in Figure \ref{fig:compentflat}).  Note that the initial value of $S$ is zero and that the M2M model may produce a negative final value of $S$.  This is interpreted as the maximum value of $S$ that the method is able to generate given all the other constraints.

\subsection{Particle initial conditions}\label{sec:pics}
\begin{figure}
	\includegraphics[width=84mm]{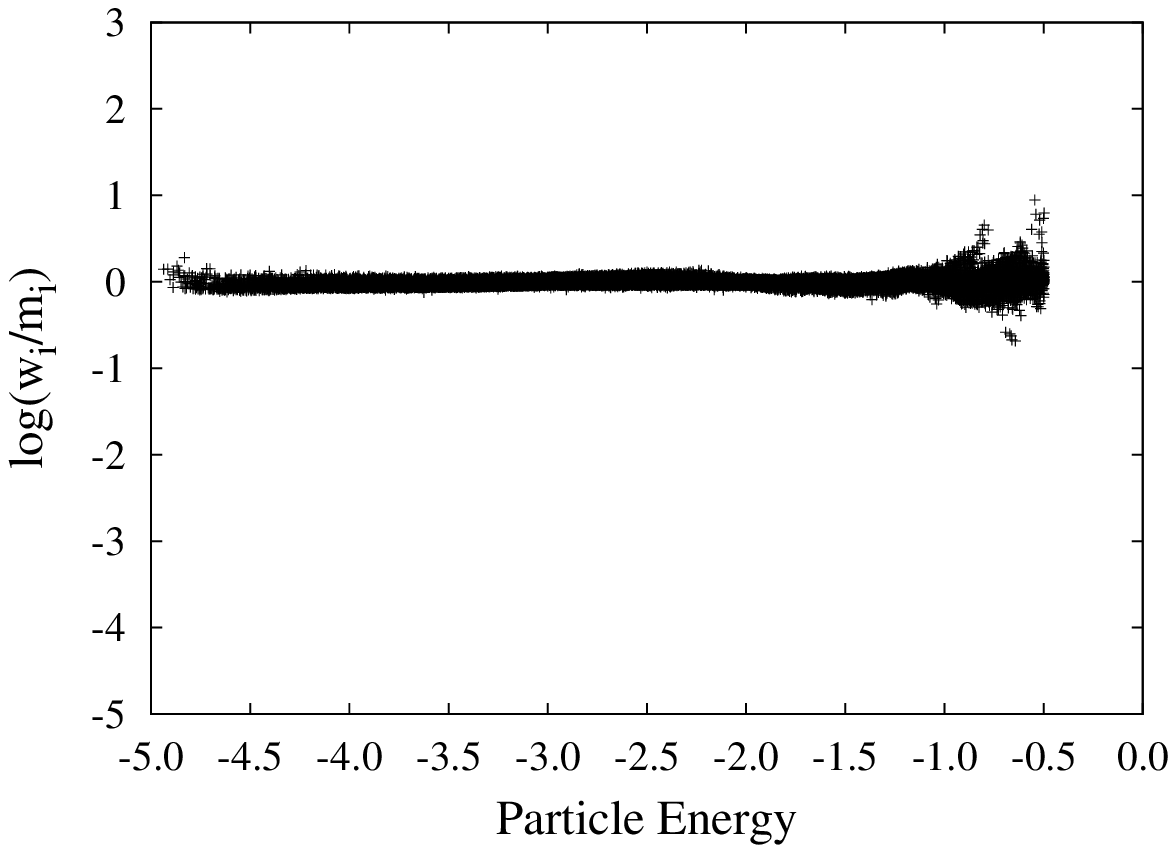}
	\includegraphics[width=84mm]{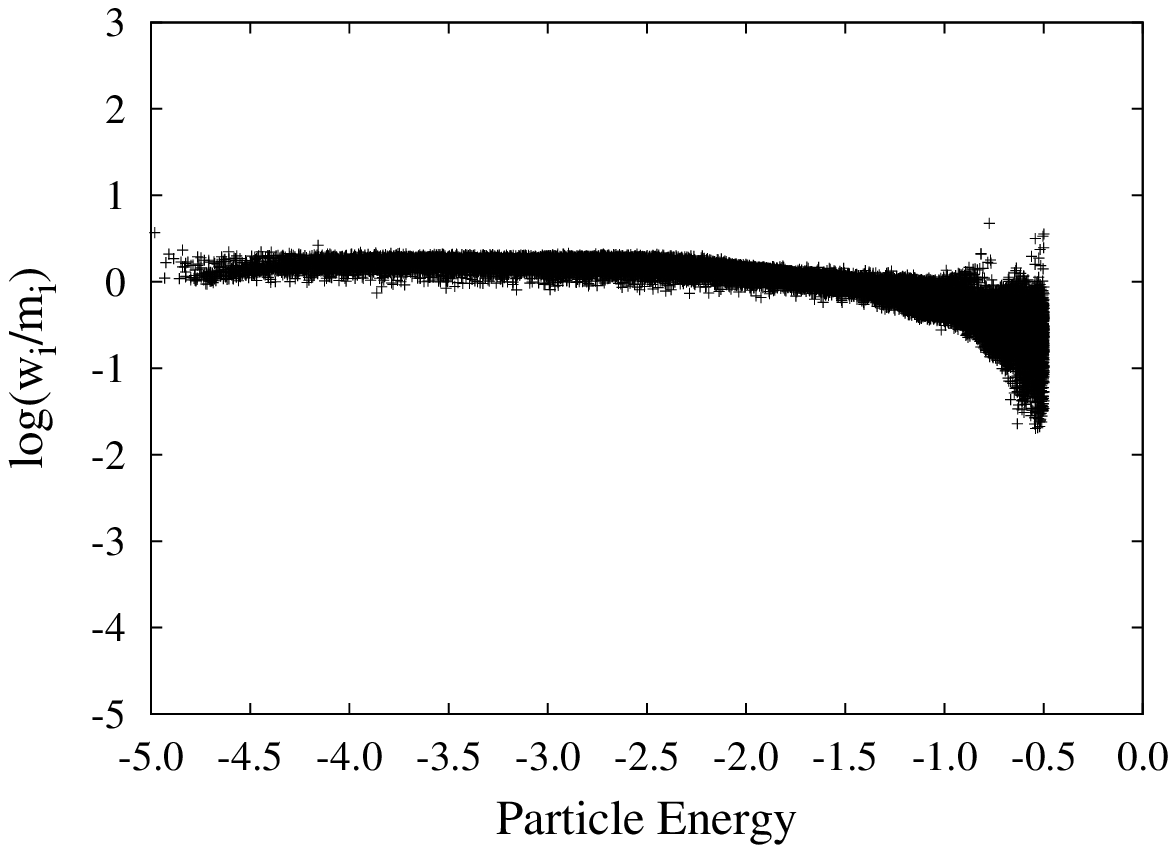}\\
\caption{End of run weight evolution comparison for the \textit{energy}-based (top panel) and \textit{random} velocity (bottom panel) schemes for creating particle initial conditions.  The weights for the high energy particles in the \textit{random} velocity scheme evolve furthest from their start values.}
\label{fig:comparespread}
\end{figure}

\begin{figure}
	\includegraphics[width=84mm]{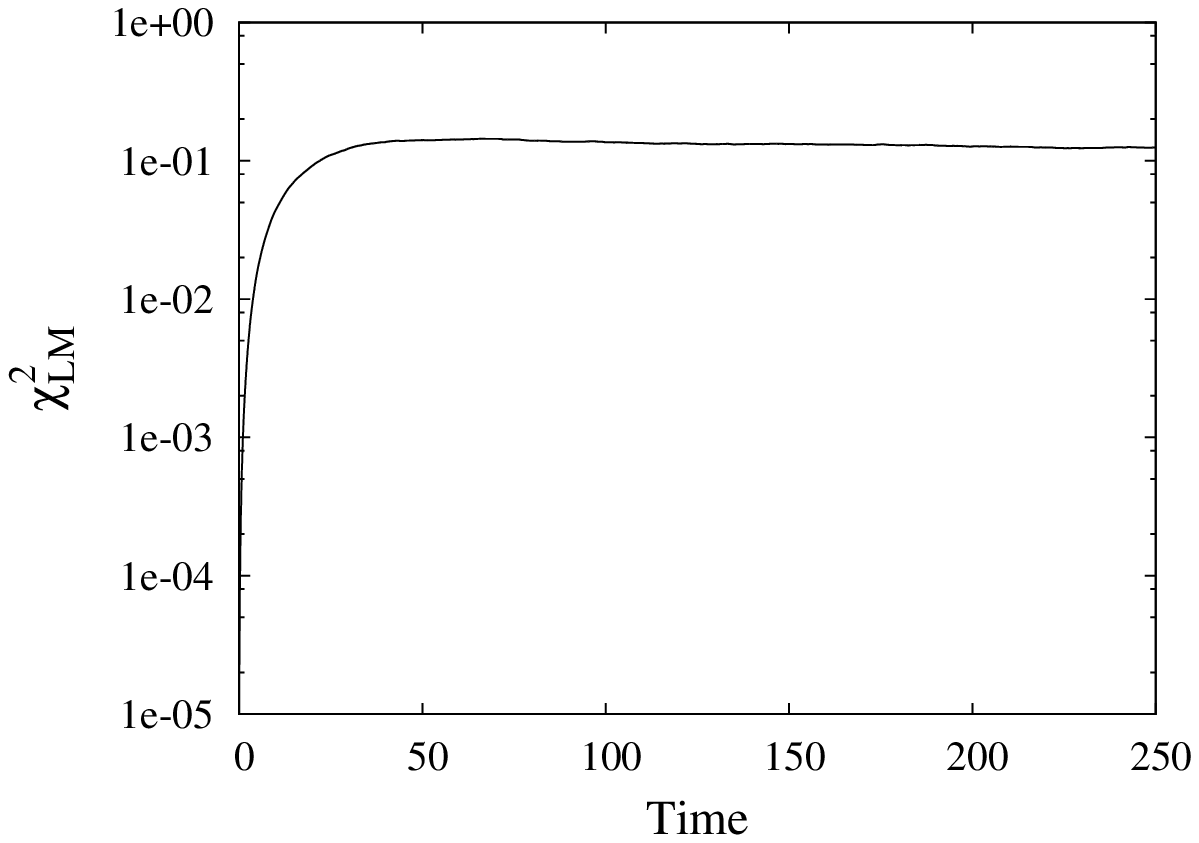}
	\includegraphics[width=84mm]{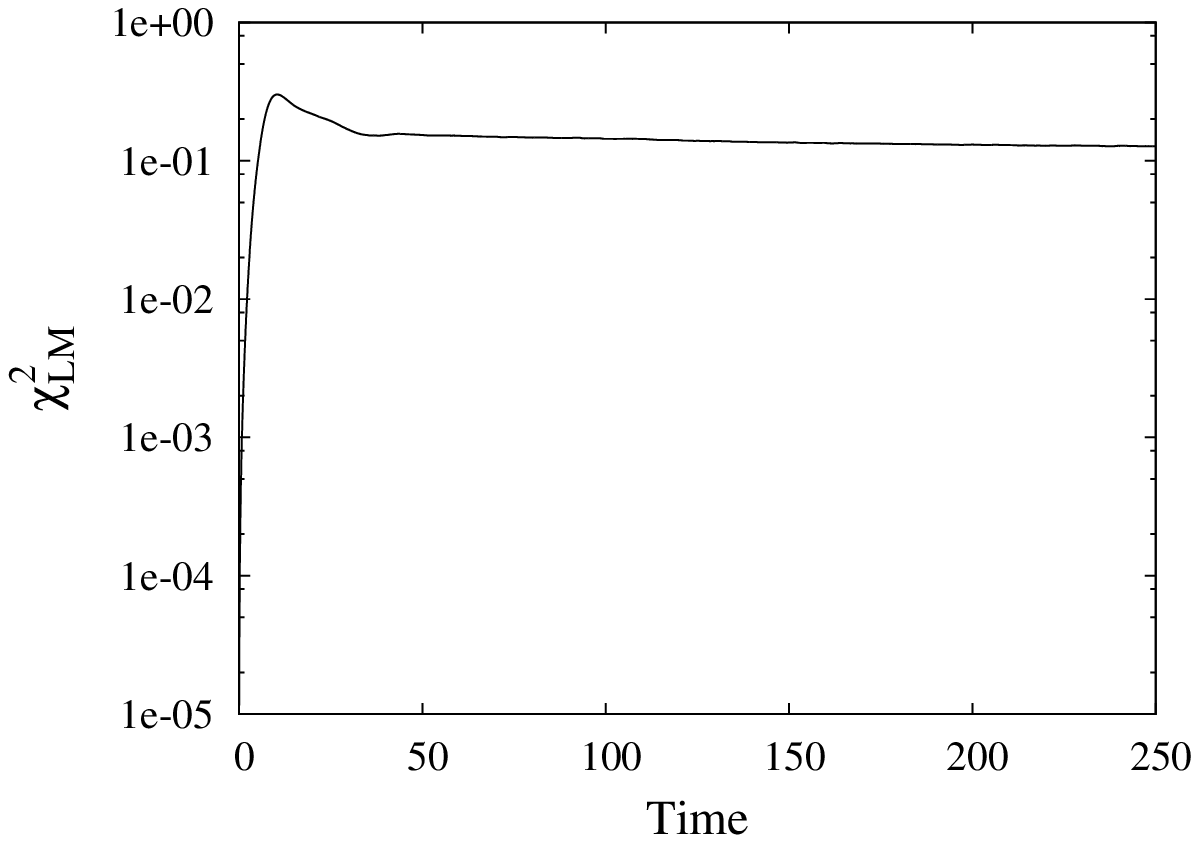}\\
	\caption{Time evolution of $\chi^2 _{\rmn{LM}}$ for the \textit{energy}-based (top panel) and \textit{Gaussian} velocity (bottom panel) schemes for creating particle initial conditions.}
	\label{fig:comparechi2}
\end{figure}

We examine the effect of the three different particle initial spatial and velocity conditions using a Plummer model with $2 \times 10^5$ particles run for $250$ time units.  The full parameter set is recorded in Table \ref{tab:icmodel} and the results are shown in Table \ref{tab:ics}.  All three schemes perform satisfactorily and, as might be expected, it is the distribution function \textit{energy}-based scheme (section \ref{sec:particleics}) which yields the best results.  From Figure \ref{fig:comparespread}, it is clear that, for the \textit{random} velocity scheme, it is the high energy particles whose weights evolve furthest from their start values.  The evolution over time of the model $\chi^2 _{\rmn{LM}}$ values (Figure \ref{fig:comparechi2}) shows differences between the schemes.  In particular, the \textit{energy}-based $\chi^2 _{\rmn{LM}}$ shows no initial `overshoot' as the particles start orbiting.
\begin{table}
	\begin{center}
	\caption{Particle initial conditions modelling parameters}
	\label{tab:icmodel}
	\begin{tabular}{|c|c|}
			\hline
			\textbf{Parameter} & \textbf{Model Value} \\
			\hline
			Overall model size & $10$ units\\
			Number of particles & $2 \times 10^5$\\
			Model duration & $250$ units\\
			Orbit integration time step & $0.02$ units\\
			Weight convergence monitoring & $25$ units\\
			Weight convergence tolerance & $5\%$\\
			$\epsilon$ & $2.5 \times 10^{-3}$\\
			$\alpha$ & $5.0 \times 10^{-2}$\\
			$\mu$ & $1.0$\\
			Surface brightness, $\lambda _{\rmn{SB}}$  & $5.0 \times 10^{-4}$ \\
			Second velocity moment, $\lambda _{\rmn{V2}}$  & $5.0 \times 10^{-3}$ \\
			Gauss-Hermite $h_4$, $\lambda _{\rmn{h4}}$  &  $2.0 \times 10^{-3}$\\
			Sum of weights, $\lambda _{\rmn{sum}}$ & $7.0 \times 10^2$\\
			Isotropic dispersion, $\lambda _{\rmn{iso}}$ & $1.0 \times 10^{-1}$\\
			\hline
	\end{tabular}
	\end{center}
\medskip
Spatial distances are given in units of the projected half light radius and times or durations in units of the half mass dynamical time.  $32$ bins are used for the surface brightness constraint and $24$ for the velocity related constraints.
\end{table}
\begin{table*}
	\begin{center}
	\caption{Comparison of different particle initial conditions}
	\label{tab:ics}
	\begin{tabular}{ccccccccc}
		\hline
		\textbf{Run} & $\bmath{-F}$ & $\bmath{\chi ^2 _{\rmn{LM}}}$ & $\bmath{\chi ^ 2 _{\rmn{SB}}}$ & $\bmath{\chi ^ 2 _{\rmn{V2}}}$ & \textbf{Particles} &\textbf{Uconv} & $\bmath{C _{\rmn{SB}}}$ & $\bmath{C _{\rmn{V2}}}$ \\
	& & &  &  & \textbf{Conv (\%)} & \textbf{Weight (\%)} \\
	\hline
	Energy   & $6.60 \: 10^{-2}$ & $1.23 \: 10^{-1}$ & $2.20 \: 10^1$ & $2.20 \: 10^1$ & $98.25$ & $1.96$ & $2.54 \: 10^{-2}$ & $7.92 \: 10^{-2}$\\
	Gaussian & $8.25 \: 10^{-2}$ & $1.26 \: 10^{-1}$ & $2.47 \: 10^1$ & $2.23 \: 10^1$ & $97.04$ & $1.96$ & $2.87 \: 10^{-2}$ & $1.11 \: 10^{-1}$\\
	Random   & $2.09 \: 10^{-1}$ & $1.47 \: 10^{-1}$ & $5.35 \: 10^1$ & $2.30 \: 10^1$ & $94.86$ & $2.05$ & $1.43 \: 10^{-1}$ & $1.64 \: 10^{-1}$\\
	\hline
	\end{tabular}
	\end{center}
\medskip
\textit{Random} indicates that the spatial distribution matches the luminosity density with random velocity components while \textit{Gaussian} has the same spatial distribution as \textit{Random} but velocity components calculated by Gaussian sampling using the velocity dispersion.  \textit{Energy} indicates that the particle energies have been calculated utilising the distribution function.
\end{table*}

\subsection{Number of particles}
We compare the effect of running a M2M model with different numbers of particles, from $10^4$ to $10^6$, and show the results in Table \ref{tab:numpart}.  We use \textit{energy}-based particle initial conditions and the same parameter settings as in section \ref{sec:pics}.
\begin{table*}
	\centering
	\caption{Impact of increasing the number of particles}
	\label{tab:numpart}
	\begin{tabular}{ccccccccc}
		\hline
		\textbf{Number} & $\bmath{-F}$ & $\bmath{\chi ^2 _{\rmn{LM}}}$ & $\bmath{\chi ^ 2 _{\rmn{SB}}}$ & $\bmath{\chi ^ 2 _{\rmn{V2}}}$ & \textbf{Particles} &\textbf{Uconv} & $\bmath{C _{\rmn{SB}}}$ & $\bmath{C _{\rmn{V2}}}$ \\
	\textbf{Particles}& & &  &  & \textbf{Conv (\%)} & \textbf{Weight (\%)} \\
		\hline
		$1 \: 10^4$ & $7.20 \: 10^{-2}$ & $1.29 \: 10^{-1}$ & $3.01 \: 10^{1}$ & $2.15 \: 10^{1}$ & $95.93$ & $4.50$ & $2.36 \: 10^{-2}$ & $1.18 \: 10^{-1}$\\
		$2 \: 10^4$ & $6.59 \: 10^{-2}$ & $1.21 \: 10^{-1}$ & $2.59 \: 10^{1}$ & $2.07 \: 10^{1}$ & $97.58$ & $2.67$ & $2.85 \: 10^{-2}$ & $9.59 \: 10^{-2}$\\
		$5 \: 10^4$ & $6.31 \: 10^{-2}$ & $1.18 \: 10^{-1}$ & $2.52 \: 10^{1}$ & $2.03 \: 10^{1}$ & $98.76$ & $1.41$ & $2.51 \: 10^{-2}$ & $1.20 \: 10^{-1}$\\
		$1 \: 10^5$ & $6.26 \: 10^{-2}$ & $1.17 \: 10^{-1}$ & $2.55 \: 10^{1}$ & $2.02 \: 10^{1}$ & $99.20$ & $0.96$ & $2.22 \: 10^{-2}$ & $1.15 \: 10^{-1}$\\
		$2 \: 10^5$ & $6.27 \: 10^{-2}$ & $1.17 \: 10^{-1}$ & $2.51 \: 10^{1}$ & $2.04 \: 10^{1}$ & $99.44$ & $0.72$ & $2.43 \: 10^{-2}$ & $1.13 \: 10^{-1}$\\
		$5 \: 10^5$ & $6.24 \: 10^{-2}$ & $1.17 \: 10^{-1}$ & $2.52 \: 10^{1}$ & $2.04 \: 10^{1}$ & $99.54$ & $0.58$ & $2.39 \: 10^{-2}$ & $1.10 \: 10^{-1}$\\
		$1 \: 10^6$ & $6.23 \: 10^{-2}$ & $1.17 \: 10^{-1}$ & $2.57 \: 10^{1}$ & $2.03 \: 10^{1}$ & $99.59$ & $0.53$ & $2.28 \: 10^{-2}$ & $1.09 \: 10^{-1}$\\
		\hline
	\end{tabular}
	
\medskip
Increasing the number of particles increases $F$, reduces the model $\chi ^2 _{\rmn{LM}}$, increases particle weight convergence and reduces the total weight associated with particles with unconverged weights.  
\end{table*}
For the constraints used, the M2M model performs well for all numbers of particles. As might be expected, increasing the number of particles increases $F$, reduces the model $\chi ^2 _{\rmn{LM}}$, increases particle weight convergence and reduces the total weight associated with particles with unconverged weights.  

Once an acceptable level of behaviour has been achieved from a M2M model, there is little to be gained by just increasing the number of particles without altering some other aspect of the model (for example, improving the spatial coverage of the observational constraints).  Running the model with a low number of particles is attractive because of the reduced computer run times - the $10^4$ particle run in Table \ref{tab:numpart} took $\approx 2$ minutes on a 2.8 GHz `dual core' workstation.  However, the weight stability of the application (running the model with weight evolution turned off) must be considered.  The stability of a $10^4$ particle run is worse (more variation in the model outputs over time) than that of a $10^5$ or a $5 \times 10^5$ particle run.

Changing the particle initial conditions to the \textit{Gaussian} velocity scheme shows only a minor degradation in model behaviour - weight convergence, for example, reduces by less than $0.5\%$.

\section{Application - Mass-to-light Determination}\label{sec:appml}
\subsection{Overview}
We illustrate a practical application of the M2M method by using it to determine the mass-to-light ratio of a simple spherical Plummer model \citep{Plummer1911}.  Assuming that mass follows light and that the mass-to-light ratio is constant, we create a data set, comprising surface brightness, line-of-sight velocity dispersion and $h_4$ Gauss-Hermite coefficient values, with a known mass-to-light ratio ($5$ in this case).  We run our M2M implementation with different mass-to-light ratios and expect that the data mass-to-light ratio will be indicated by a minimum in the values of $-F$ and the model $\chi^2$ values.  

\subsection{Model and data preparation}\label{sec:mldataprep}
We use constructed functions for luminosity density $\nu (R, x _{\parallel})$ and velocity dispersion $\sigma (R, x _{\parallel})$ as described in section \ref{sec:finitemodel}.  In the following, we take the total luminosity as $1$, the gravitational constant $G = 1$, $\Upsilon$ as the mass-to-light ratio, $R$ as the projected radius and $r$ as the spherical radius.  Spatial distances are given in units of the projected half light radius ($=1$ for our model).  The key Plummer model expressions we require are
\begin{enumerate}
\item Surface brightness
\begin{equation}
	I(R) = \int dx_{\parallel} \nu (R, x _{\parallel}).
\end{equation}

\item Luminosity weighted line-of-sight velocity dispersion
\begin{equation}
	\sigma _{\parallel} ^2 (R) = \frac{\int dx_{\parallel} \nu (R, x _{\parallel}) \sigma ^2 (R, x _{\parallel})}{I(R)}.
\end{equation}

\item Distribution function
\begin{equation}
	f( \left \vert E \right \vert) \propto \left \vert E \right \vert ^{7/2}.
\end{equation}

\item Potential
\begin{equation}
	\phi(r) = -\frac{\Upsilon}{\left ( r^2 + 1 \right) ^{1/2}}.
\end{equation}
\end{enumerate}

The observational constraints are surface brightness, surface brightness times line-of-sight velocity dispersion squared and surface luminosity times the $h_4$ Gauss-Hermite coefficient.  Surface brightness is limited to a radial extent of $8$ units and the velocity constraints to $5$.  We bin all observables radially, with pseudo-logarithmic bin sizes \citep{Sellwood2003}, utilising $32$ bins for surface brightness and $24$ bins for the velocity constraints.  For the error terms, $\sigma(Y_j)$, we use a relative error of $5$ per cent for surface brightness and $10\%$ for line-of-sight velocity dispersion.  For $h_4$, we use an absolute error of $0.015$ which is consistent with the published SAURON rms error \citep{Sauron2}. We create the surface brightness and line-of-sight velocity dispersion data values as described in section \ref{sec:gobs} using a 2-sigma cut-off.  For $h_4$, we just take the theoretical values (times the bin luminosity). In addition to the observational constraints , we also impose the total weight and the isotropic velocity dispersion constraints.

For the purposes of this mass-to-light illustration, the particle initial conditions are \textit{energy} based as described in section \ref{sec:particleics} and are created using the mass-to-light ratio of the model run, not the ratio with which the data was created.  The particle initial weights and priors are set as $1/N$ where $N$ is the number of particles.

\subsection{Results}\label{sec:mlresults}
\begin{figure}
	\includegraphics[width=84mm]{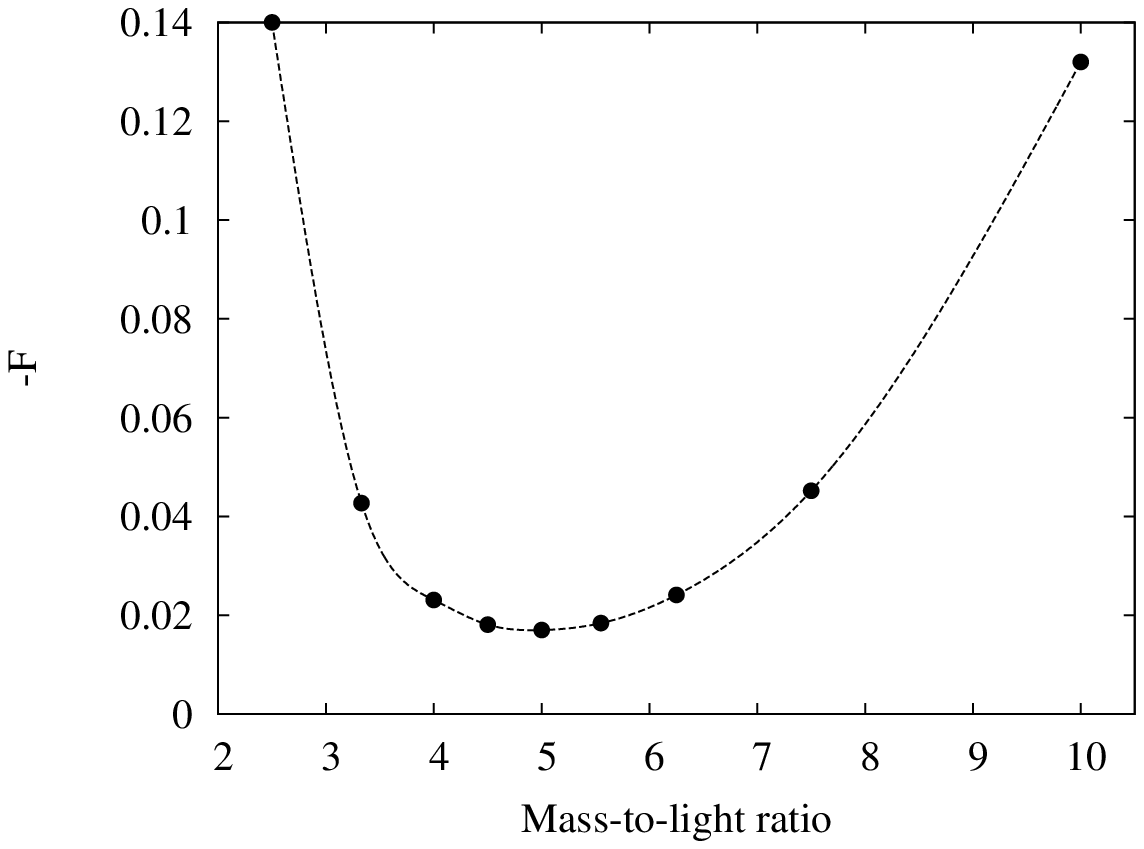}
	\includegraphics[width=84mm]{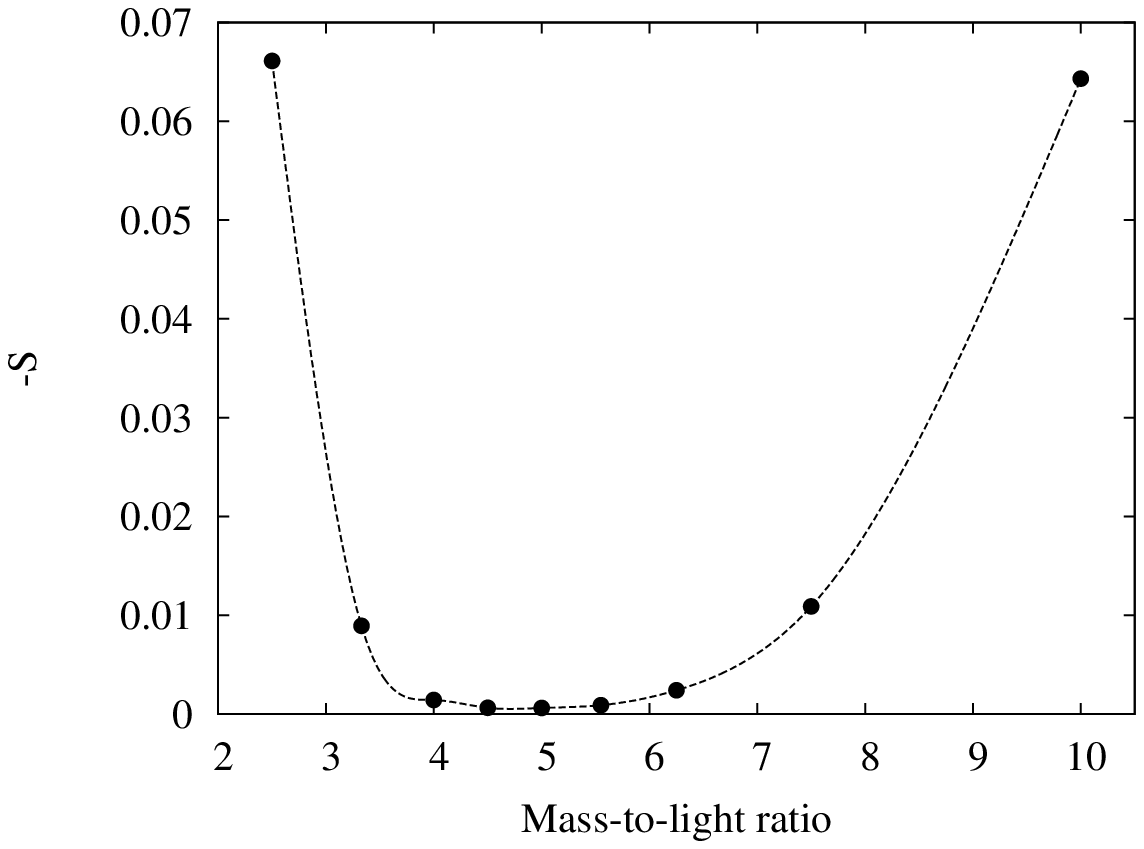}
	\includegraphics[width=84mm]{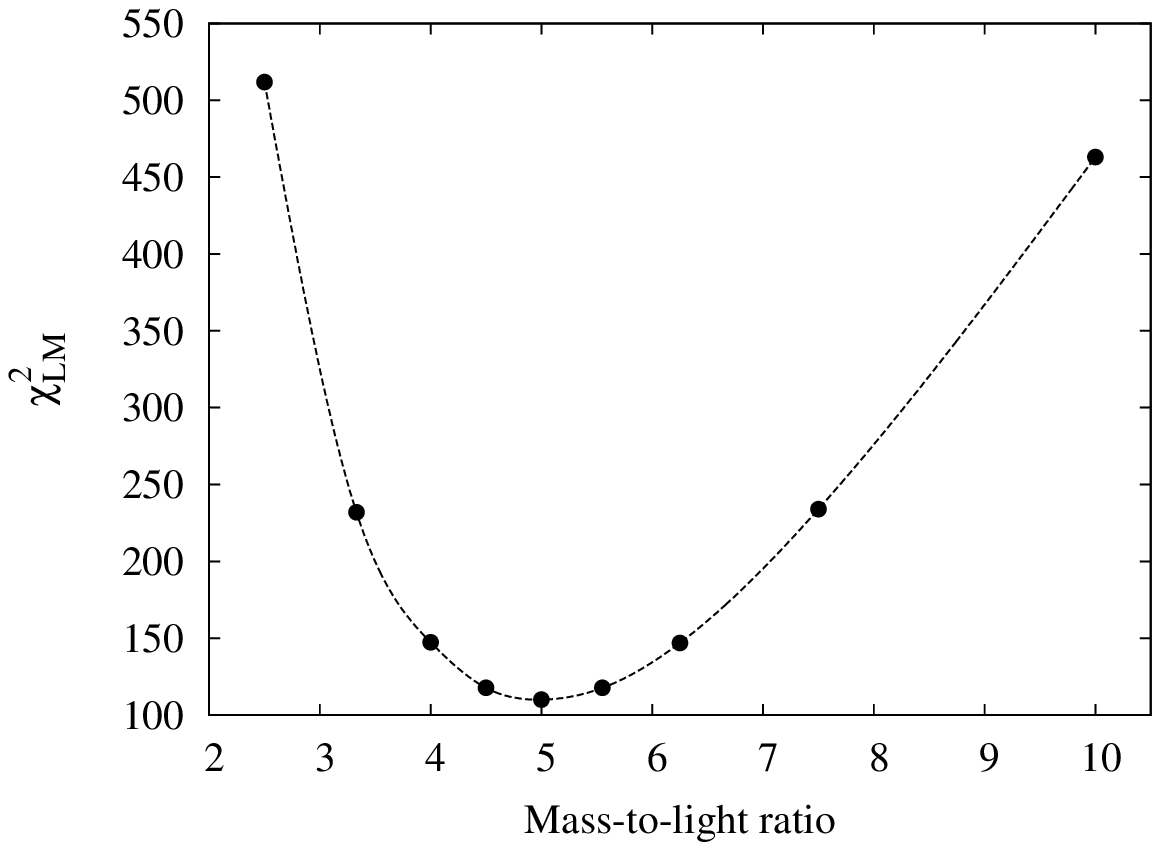}
	\caption{End of run $-F$, $-S$ and $\chi ^2 _{\rmn{LM}}$ plotted against mass-to-light ratio $\Upsilon$. The minimum values are at $\Upsilon = 4.91$, $\Upsilon = 4.71$ and $\Upsilon = 4.97$ respectively.}
	\label{fig:m2mplot}
\end{figure}

\begin{figure}
	\includegraphics[width=84mm]{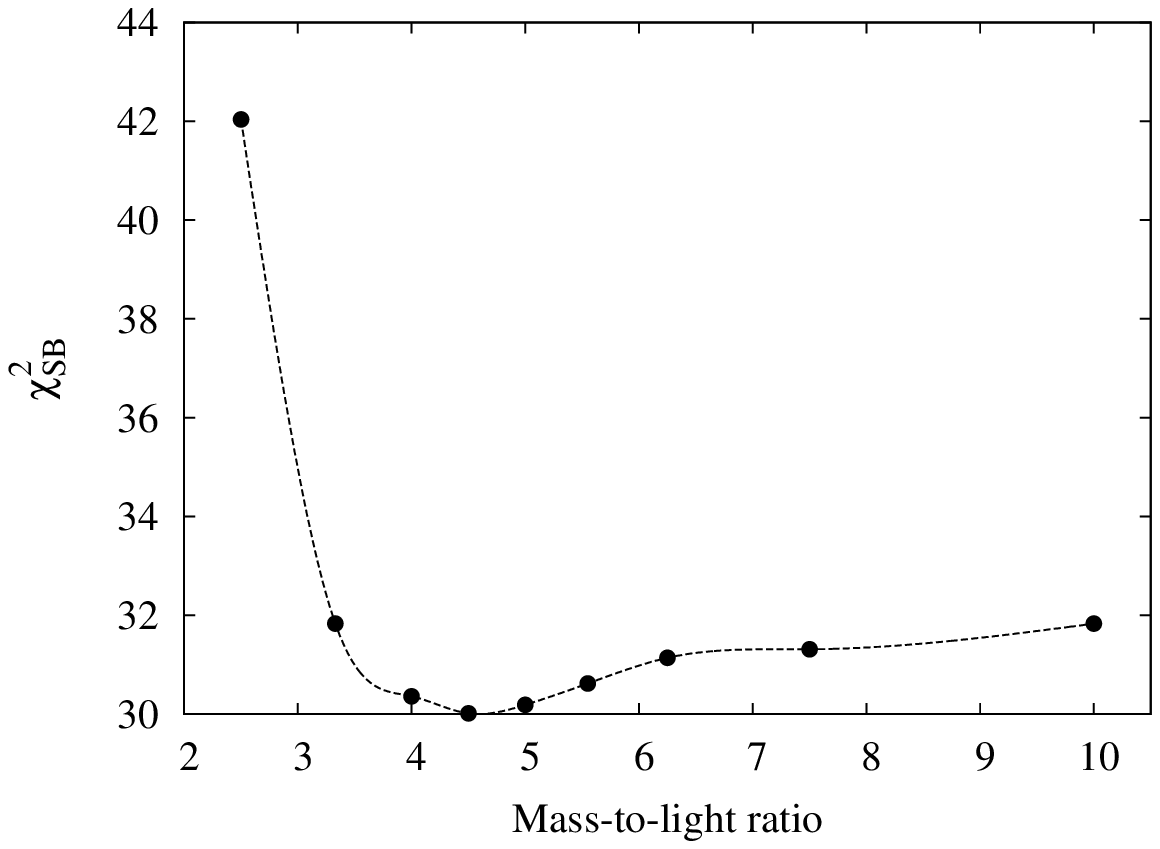}
	\includegraphics[width=84mm]{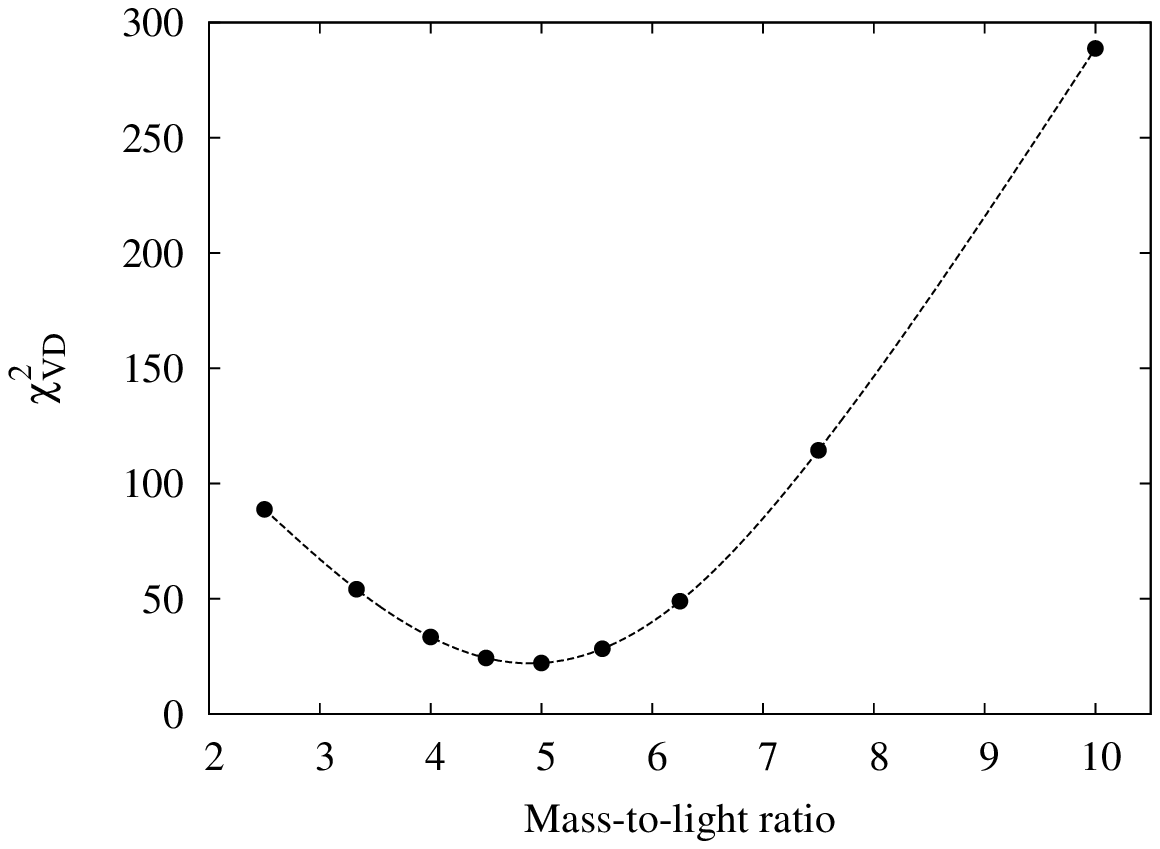}
	\includegraphics[width=84mm]{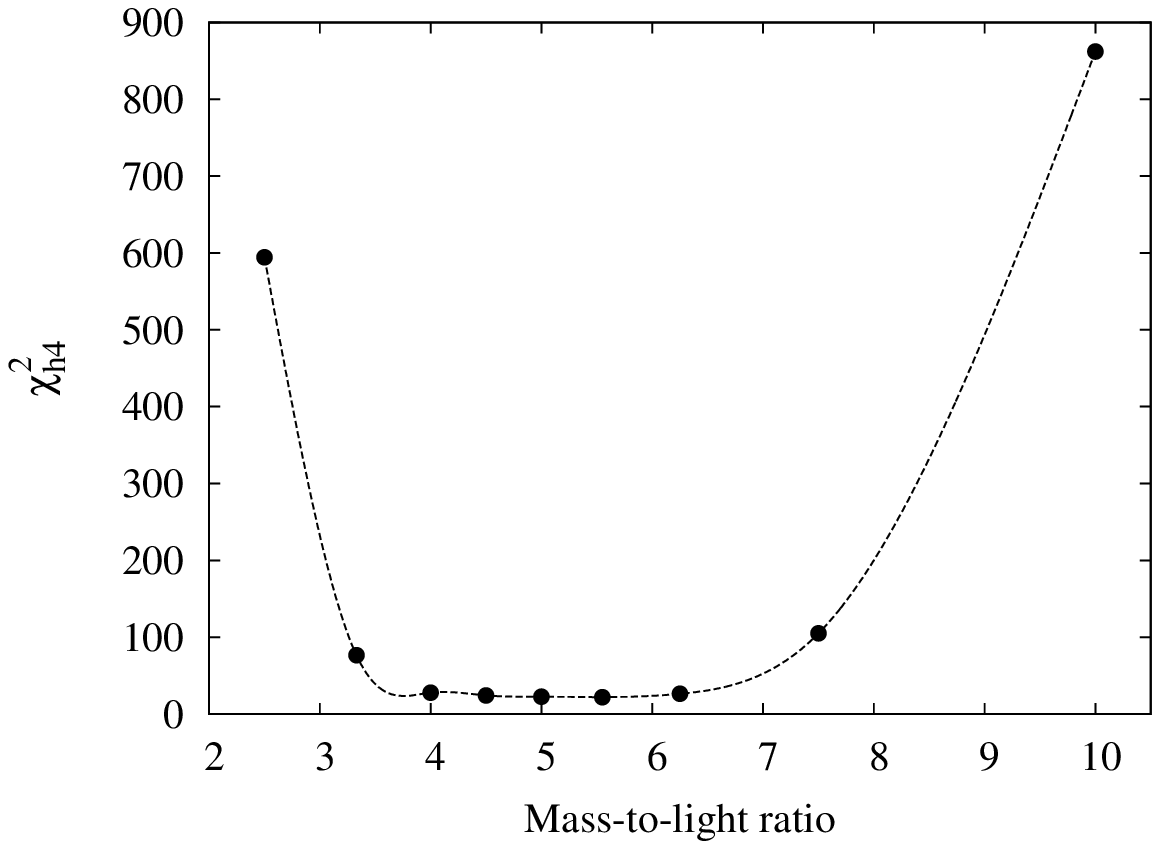}
	\caption{End of run $\chi^2 _{\rmn{SB}}$, $\chi^2 _{\rmn{VD}}$ and $\chi^2 _{\rmn{h4}}$ plotted against mass-to-light ratio $\Upsilon$.  The minimum values occur at $\Upsilon = 4.59$, $\Upsilon = 4.90$ and $\Upsilon = 5.52$ respectively.}
	\label{fig:indivchi2}
\end{figure}

\begin{table}
	\begin{center}
	\caption{Mass-to-light modelling parameters}
	\label{tab:mlmodel}
	\begin{tabular}{|c|c|}
			\hline
			\textbf{Parameter} & \textbf{Model Value} \\
			\hline
			Data mass-to-light ratio & $5$\\
			Overall model size & $10$ units\\
			Number of particles & $5 \times 10^4$\\
			Model duration & $250$ units\\
			Weight convergence monitoring & $25$ units\\
			Weight convergence tolerance & $5\%$\\
			$\epsilon$ & $2.5 \times 10^{-3}$\\
			$\alpha$ & $5.0 \times 10^{-2}$\\
			$\mu$ & $1.0$\\
			$\lambda _{\rmn{SB}}$  & $5.0 \times 10^{-4}$ \\
			$\lambda _{\rmn{VD}}$  & $2.5 \times 10^{-3} / \Upsilon$ \\
			$\lambda _{\rmn{h4}}$  &  $2.0 \times 10^{-3}$\\
			$\lambda _{\rmn{sum}}$ & $10^3$\\
			$\lambda _{\rmn{iso}}$ & $5.0 \times 10^{-1}$\\
			Line of sight axis & x axis\\
			\hline
	\end{tabular}
	\end{center}
\medskip
Spatial distances are given in units of the projected half light radius and times or durations in units of the half mass dynamical time.
\end{table}
We execute the M2M model, using the parameters in Table \ref{tab:mlmodel}, for 9 different mass-to-light ratios and plot the  resulting model $-F$, $-S$ and $\chi^2 _{\rmn{LM}}$ values against mass-to-light ratio.  As can be seen from Figure \ref{fig:m2mplot}, these model values have a minimum at a mass-to-light ratio of $\approx 5$.
We fit smooth curves to $-F$, $-S$ and  $\chi ^2 _{\rmn{LM}}$ using cubic spline interpolation and determine that the minimum values occur at mass-to-light ratios of $4.91$, $4.71$ and $4.97$ respectively.  By removing the $\chi^2 _{\rmn{LM}}$ factors (the $\lambda _k$ in equation \ref{eqn:chifactor}) and rescaling the $\chi ^2$ curve such that the $\chi ^2$ minimum value is equal to the number of degrees of freedom ($78$ in this case), we establish the $1 \sigma$ error bounds and give the model determined value of the mass-to-light ratio of the data set as $4.88 \pm 0.21$. To complete the $\chi ^2$ analysis, we show the individual observable $\chi ^2$ plots in Figure \ref{fig:indivchi2}.

\begin{table}
	\begin{center}
	\caption{Weight convergence and dispersion isotropy}
	\label{tab:mltable}
	\begin{tabular}{cccccc}
	\hline
	$\bmath{\Upsilon}$ & \textbf{Particles} & \textbf{Unconv} & $\bmath{\sum w_i}$ & $\bmath{\beta (0)}$ & $\bmath{\beta}$ \\
		          & \textbf{Converged} & \textbf{Weight} & & & \textbf{grad} \\
		          & \textbf{(\%)} & \textbf{(\%)}\\
	\hline
	$2.50$ & $93.5$ & $6.5$ & $1.0$ & $+0.001$ & $-0.026$ \\
	$3.33$ & $98.3$ & $1.7$ & $1.0$ & $-0.003$ & $-0.011$ \\
	$4.00$ & $98.3$ & $1.6$ & $1.0$ & $-0.003$ & $-0.006$ \\
	$4.50$ & $99.1$ & $0.9$ & $1.0$ & $-0.009$ & $-0.002$ \\
	$5.00$ & $98.7$ & $1.3$ & $1.0$ & $-0.016$ & $+0.003$ \\
	$5.55$ & $98.3$ & $1.6$ & $1.0$ & $-0.014$ & $+0.002$ \\
	$6.25$ & $98.3$ & $1.6$ & $1.0$ & $-0.014$ & $+0.002$ \\
	$7.50$ & $98.3$ & $1.6$ & $1.0$ & $-0.017$ & $+0.006$ \\
	$10.0$ & $97.8$ & $2.0$ & $1.0$ & $-0.020$ & $+0.006$ \\
	\hline
	\end{tabular}
	\end{center}

\medskip
Weight convergence peaks close to the true mass-to-light ratio.  The total weight constraint is met for all runs and the velocity dispersion is isotropic except for $\Upsilon = 2.5$.
\end{table}

The total weight constraint is met (Table \ref{tab:mltable}) and so is the velocity dispersion isotropy constraint except at $\Upsilon = 2.5$.  Weight convergence is high, peaking close to the true mass-to-light ratio, and, no less important, the weight associated with particles with unconverged weights is low at $<2\%$ of the total particle weight. 

\begin{figure}
	\centering
	\includegraphics[width=84mm]{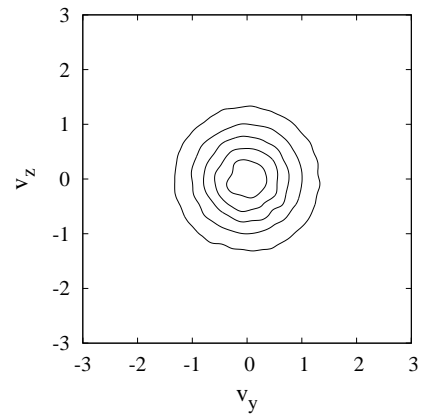}
	\caption{Particle weight distribution in velocity space perpendicular to the line of sight for $\Upsilon = 5$.  The highest valued contours are in the centre with the lowest on the outside.}
	\label{fig:ml5vyvz}
\end{figure}

\begin{figure}
	\centering
	\includegraphics[width=84mm]{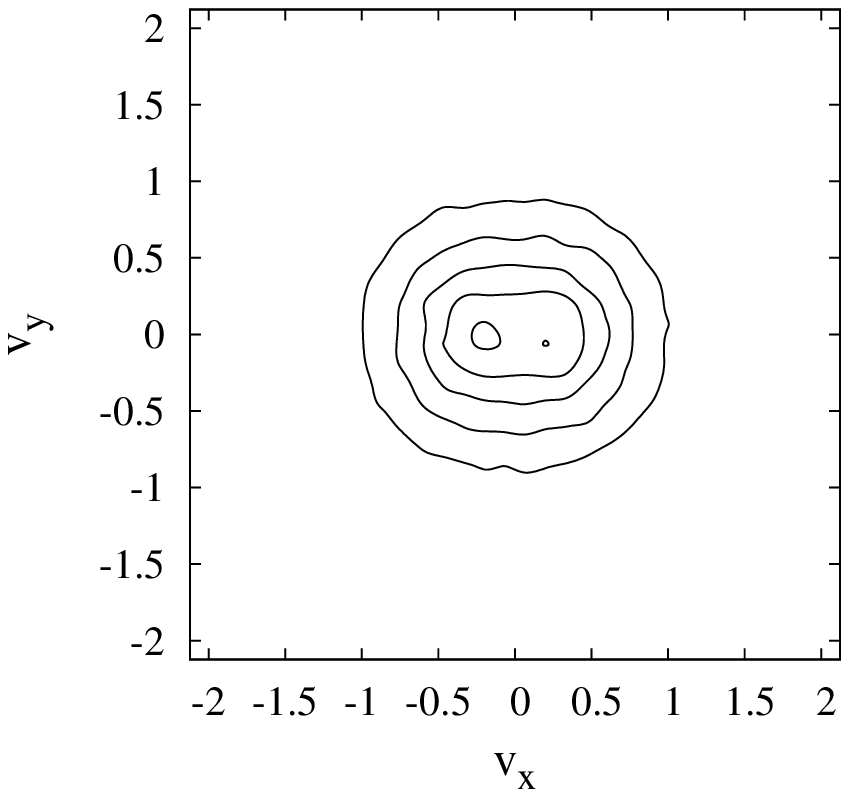}
	\includegraphics[width=84mm]{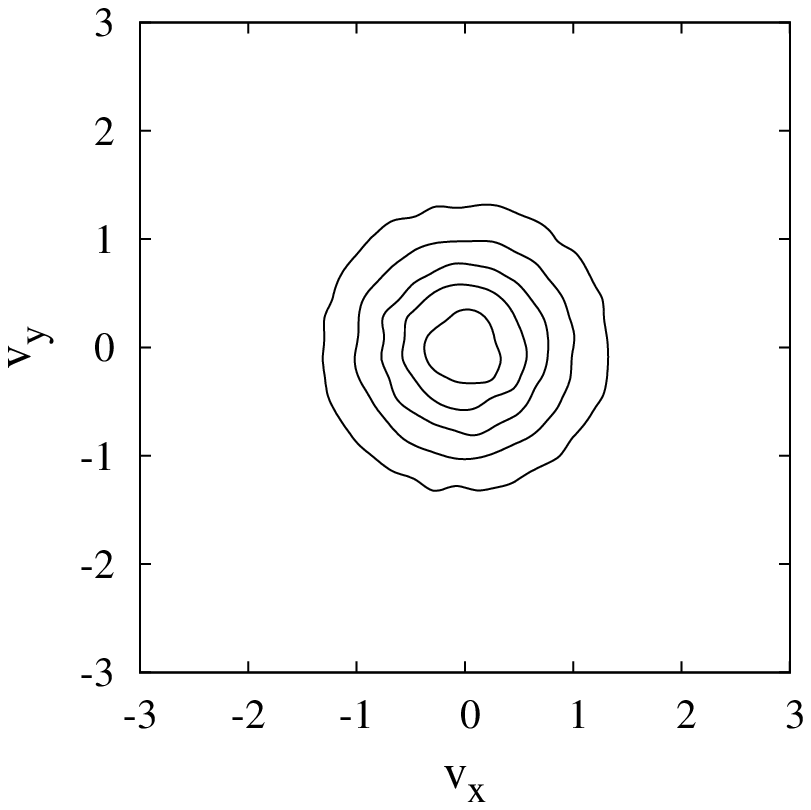}
	\includegraphics[width=84mm]{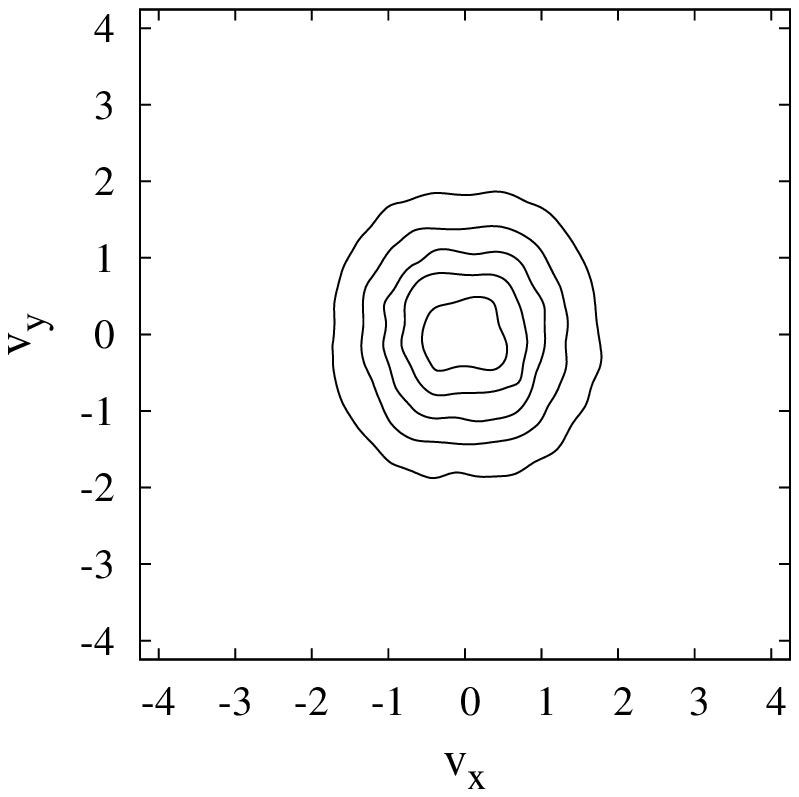}\\
	\caption{Particle weight distribution in velocity space parallel to the line of sight for $\Upsilon = 2.5$ (top), $\Upsilon = 5$ (target, centre) and $\Upsilon = 10$ (bottom). As described in section \ref{sec:mlresults}, the contours show stretching and compression in the $v_x$ direction as the M2M model tries to match the supplied velocity dispersion constraint for $\Upsilon$ values less than and greater than the target value.  The reason for the double peak in the $\Upsilon = 2.5$ plot is not known.}
	\label{fig:vxvyplots}
\end{figure}

By examining the particle weight distribution in velocity space, we investigate how the M2M method behaves given a mass-to-light value which does not match the observational data set.  For our Plummer model with the x-axis orientated to the line of sight, plotting the weight distribution contours perpendicular to the line of sight (in the $v_z - v_y$ plane) we would expect to see a circular pattern as in Figure \ref{fig:ml5vyvz} and also in the other two planes.  The $\Upsilon = 5$ $v_x - v_y$ plot (Figure \ref{fig:vxvyplots}) is as expected. 
However, the $\Upsilon = 2.5$ plot is stretched in the $v_x$ direction whilst the $\Upsilon = 10$ plot is compressed.  Given the maximum velocity for $\Upsilon = 2.5$ is less than that of the data set ($\Upsilon = 5$), it is reasonable to expect that the M2M model selectively increases the weights of particles to try and reproduce the data set line-of-sight velocity dispersion.  Similarly for  $\Upsilon = 10$, with a greater maximum velocity, the weights of particles are selectively reduced.  The end result is the distorted particle weight distributions.  

\begin{figure*}
	\includegraphics[width=84mm]{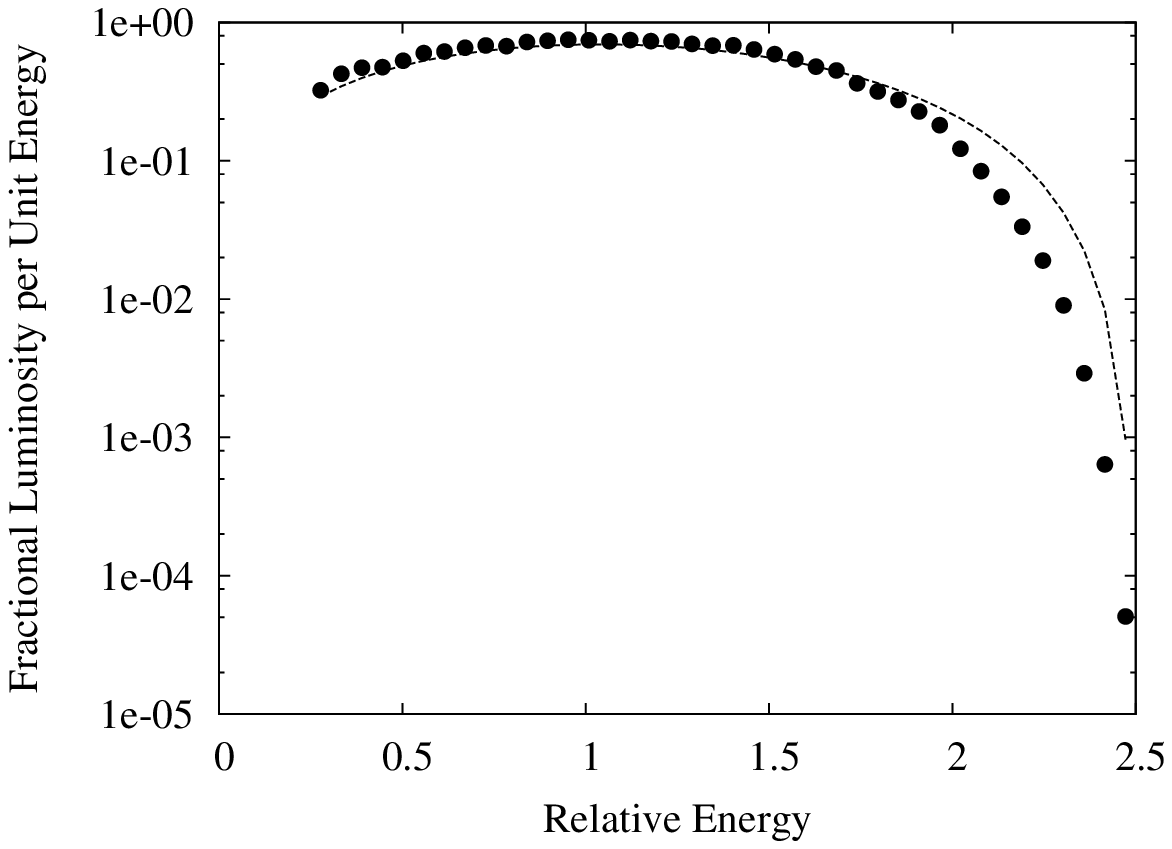}
	\includegraphics[width=84mm]{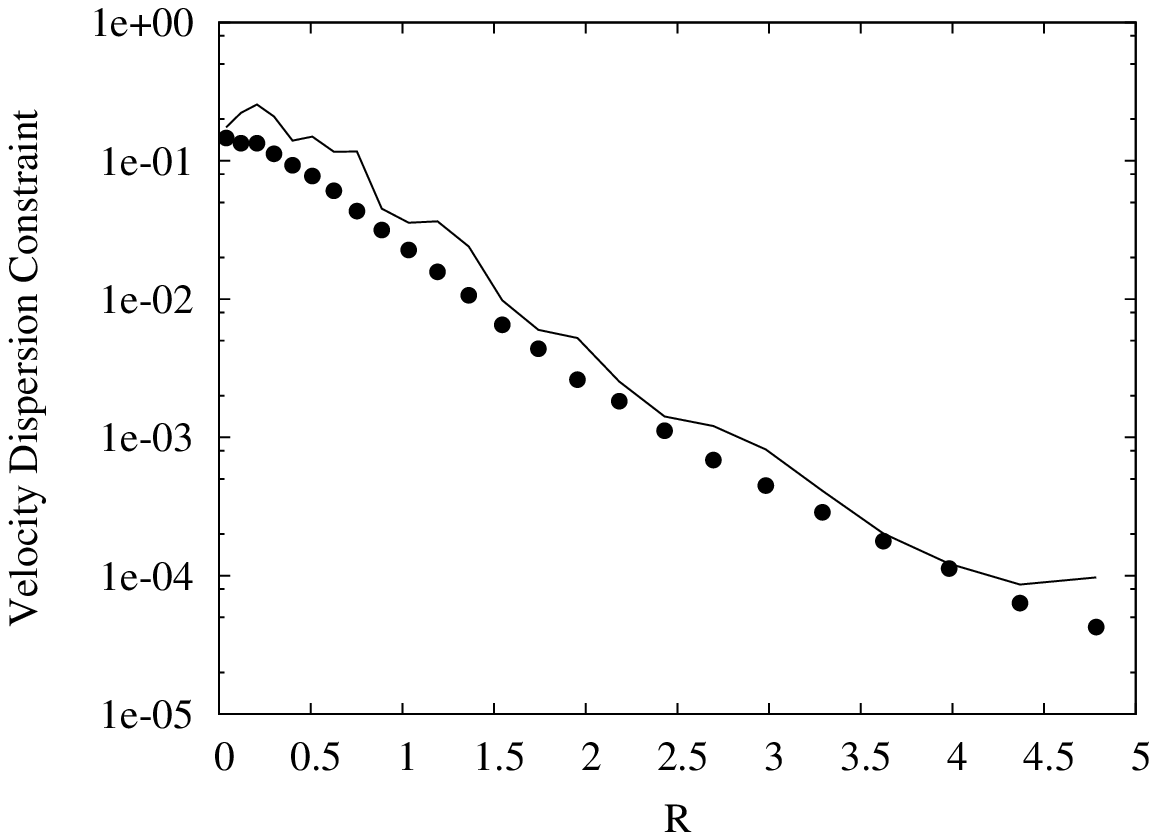}\\
	\includegraphics[width=84mm]{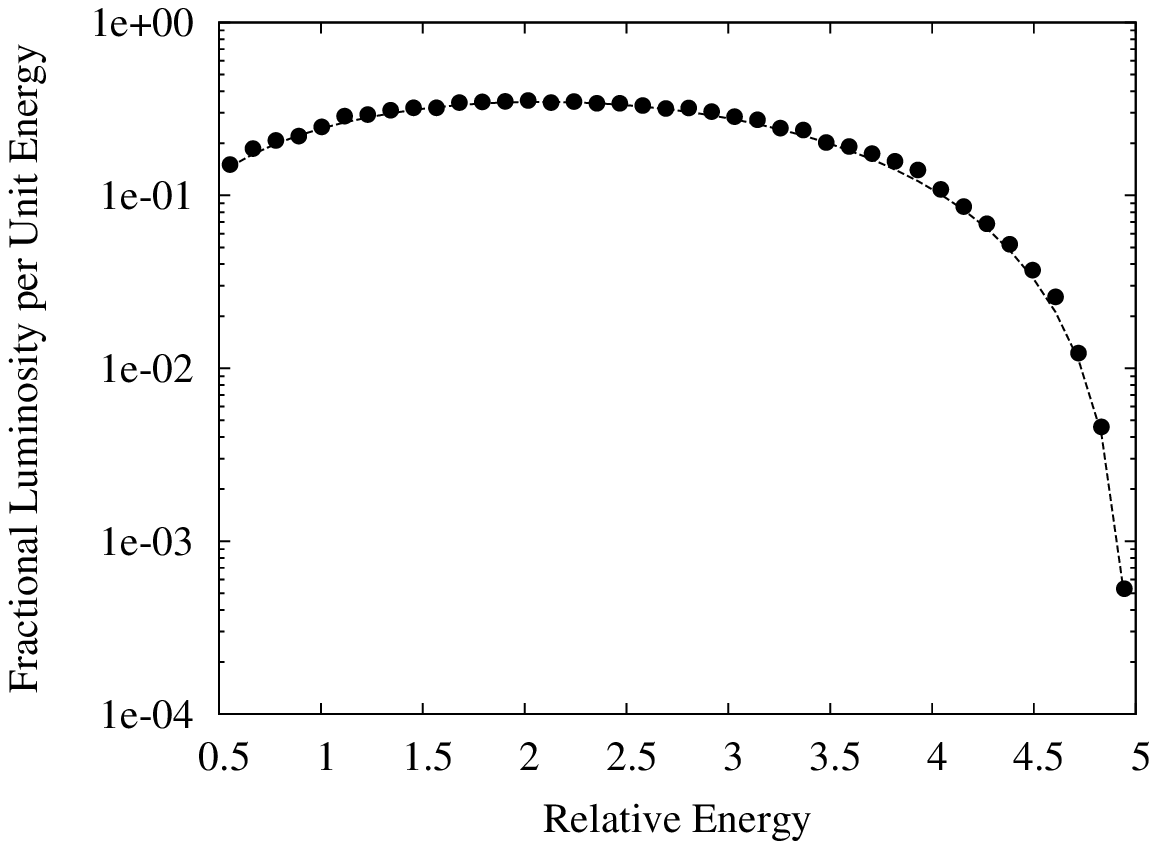}
	\includegraphics[width=84mm]{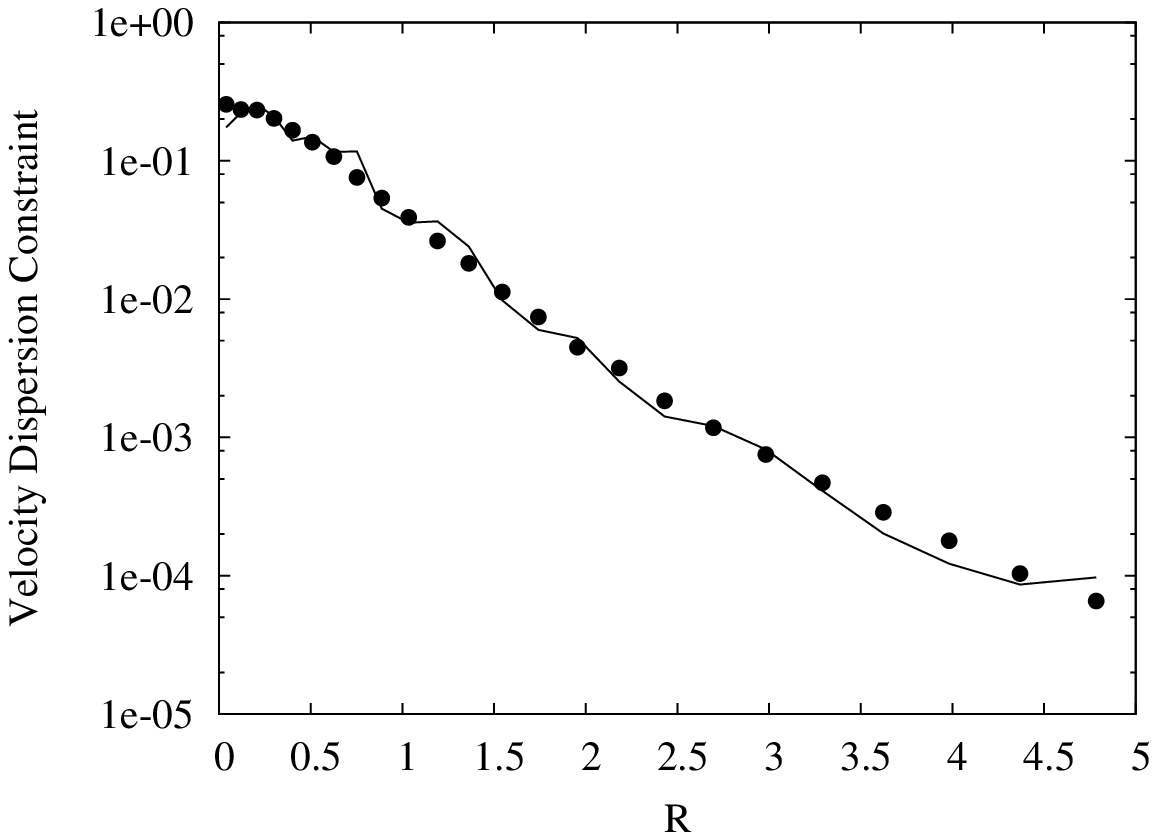}\\
	\caption{Reproduction of $dL/dE$ and velocity dispersion.  The rows, from the top, are for $\Upsilon = 2.5$ and $\Upsilon = 5$.  The solid circles are the model produced values. In the left hand panels, the dotted line is the theoretical function.  In the right hand panels, the solid line indicates the observable constraint data without error terms. The dispersion constraint is not met for $\Upsilon = 2.5$.   The M2M model is unable to increase the particle weights sufficiently to reproduce the constraint. The $dL/dE$ plots show that is the high relative energy particles which are most obviously affected with the lower energy particle weights being adjusted to compensate. The converse arguments apply for mass-to-light ratios higher than the true ratio.}
	\label{fig:reprovd2.5}
\end{figure*}

In Figure \ref{fig:reprovd2.5}, we display plots for different mass-to-light ratios showing the model luminosity per unit energy $dL/dE$ compared with the theoretical function and the velocity dispersion constraint compared with the observable data.  At lower mass-to-light ratios, calculating $dL/dE$ from the end-of-run particle table shows that it is increased at low energies and reduced at high energies by comparison with its initial values. The converse is true for higher mass-to-light ratios with $dL/dE$ being reduced at low energies and increased at high. 

\subsection{Different constraints and initial conditions}\label{sec:dcic}
Clearly there are other combinations of constraining observables and particle initial conditions we could have used.  We now examine the effect of replacing surface brightness (run 9 in Table \ref{tab:dcictable}) by luminosity density as a constraint (run 12), using an alternative velocity dispersion constraint (run 14), changing the particle initial conditions from \textit{energy} based to \textit{Gaussian} velocity (run 15) and \textit{random} velocity (run 16), and finally increasing the number of particles from $5 \times 10^4$ to $2 \times 10^5$ (run 10).  For the alternative velocity dispersion constraint, we remove the surface brightness multiplier so that the constraint is just line-of-sight velocity dispersion squared.
\begin{table}
	\begin{center}
	\caption{Different constraints and initial conditions}
	\label{tab:dcictable}
	\begin{tabular}{cccc}
	\hline
	\textbf{Run} & $\bmath{\chi ^2 _{\rmn{LM}}}$ & $\bmath{\chi ^2}$ & $\bmath{\chi ^2 _{\rmn{VD}}}$ \\
	\hline
	$9$  & $4.97$ & $4.88 \pm 0.21$ & $4.90 \pm 0.27$  \\
	$12$ & $4.96$ & $4.88 \pm 0.18$ & $4.91 \pm 0.26$  \\
	$14$ & $5.61$ & $5.37 \pm 0.28$ & $5.38 \pm 0.33$  \\
	$15$ & $4.99$ & $4.92 \pm 0.22$ & $4.79 \pm 0.28$  \\
	$16$ & $4.80$ & $4.65 \pm 0.27$ & $4.52 \pm 0.27$  \\
	$10$ & $5.00$ & $4.89 \pm 0.22$ & $4.90 \pm 0.27$  \\
	\hline
	\end{tabular}
	\end{center}

\medskip
Run 9 is the main mass-to-light run described in section \ref{sec:mlresults}.  Run 10 using $4$ times as many particles as run 9 shows little difference in the results.  The remaining runs are as per section \ref{sec:dcic}.
\end{table}

We find that the largest variations from the true mass-to-light ratio come from using the alternative dispersion constraint and the \textit{random} velocity initial conditions.  For the other runs, the true mass-to-light is within 1-sigma of the model estimates.  Note that increasing the number of particles makes very little difference.

\subsection{Summary}
To conclude this section, we have demonstrated the M2M method in action determining the mass-to-light ratio of a galaxy modelled by a theoretical Plummer sphere model.  All the constraining observables are simple combinations of observational measurements regularly taken of actual galaxies, that is surface brightness, line-of-sight velocity dispersion and the $h_4$ Gauss-Hermite coefficient and their associated errors.  For our constructed data generated with a mass-to-light ratio of $5$, we are able to use the M2M method in a variety of ways to estimate that ratio.  Using $\chi ^2 _{\rmn{VD}}$, our best estimate is $4.91 \pm 0.26$ with a spread of results from $4.52 \pm 0.27$ to $5.38 \pm 0.33$.

\section{Draco}\label{sec:dracomain}
\subsection{Introduction}\label{sec:dracointro}
\begin{figure}
		\centering
		\includegraphics[width=84mm]{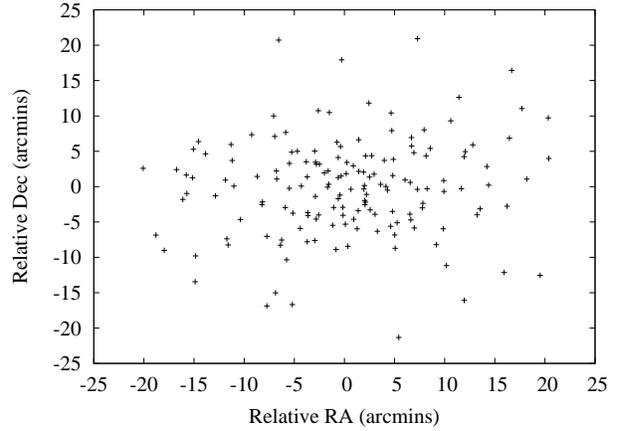}
		\caption[Spatial distribution of Draco velocity measurements]{Spatial distribution of Draco velocity measurements from \citet{Kleyna2002}.}
	\label{fig:kleynadata}
\end{figure}

Draco is a dwarf spheroidal satellite galaxy of the local group located some $75 \; \rmn{kpc}$ from the Sun and is interesting,  both cosmologically and astrophysically, in that there is no currently accepted explanation for its stellar kinematics without invoking dark matter.  \citet{Wilkinson2002} describe a mathematical model for modelling dwarf spheroidal galaxies and \citet{Kleyna2002} apply the model to Draco.  In this section, we use the M2M method with the Draco data from \citet{Kleyna2002} plus the isotropic velocity distribution model from \citet{Wilkinson2002}  to determine the mass-to-light ratio of Draco.  The data set comprises 159 line-of-sight stellar velocity measurements with their errors.  The spatial distribution of the measurements is shown in Figure \ref{fig:kleynadata}.
Binning the data to create a second velocity moment shows that the moment is rising with increasing radius.  Other key data taken from \citet{Kleyna2002} are the central velocity dispersion of $8.5\  \rmn{km} \ \rmn{s}^{-1}$, the effective radius of $9.71$ arcminutes ($\approx 214 \ \rmn{pc}$), and the central V-band surface brightness of $2.2 \times 10^6 \  \rmn{L}_{\odot}\  \rmn{kpc}^{-2}$.  Following the analysis in \citet{Wilkinson2002} and elsewhere, the surface brightness data is taken to follow a spherical Plummer model.

For an isotropic velocity dispersion with a rising line-of-sight velocity dispersion curve, the key equations based on \citet{Wilkinson2002} are
\begin{enumerate}
\item Relative potential
\begin{equation}
	\psi (r) = \frac{\psi _0}{\left ( 1 + r^2  \right ) ^{\alpha / 2}}
\end{equation}
where $\alpha < 0$.
\item Distribution function
\begin{equation}
	f(E) \propto \left |  E \right | ^{5 / \alpha - 3/2}.
\end{equation}
\item Circular velocity
\begin{equation}\label{eqn:wilkcirc}
	v ^2 _{\rmn{circ}} = v ^2 _0 \frac{r^2}{\left ( 1 + r^2  \right ) ^{1 + \alpha / 2}}
\end{equation}
where $\psi _0 = v ^2 _0 / \alpha$.  Given that the matter distribution is spherical, $v  _{\rmn{circ}}$ may also be expressed as
\begin{equation}\label{eqn:gencirc}
	v ^2 _{\rmn{circ}} = \frac{G M(<r)}{r}.
\end{equation}
\item Surface brightness
\begin{equation}
	I(R) = \frac{I _0}{\left ( 1 + R^2  \right ) ^2}
\end{equation}
where $R$ is the projected radius, and $I_0$ is the measured central surface brightness.
\item Line-of-sight second velocity moment
\begin{equation}\label{eqn:dracovd}
	\sigma ^2 (R) = \frac{\sigma ^2 _0}{\left ( 1 + R^2  \right ) ^{\alpha / 2}}
\end{equation}
where $\sigma _0$ is the measured central line-of-sight value.  The relationship between $\sigma _0$ and $v_0$ is
\begin{equation}
	\sigma ^2 _0 = \frac{3 \sqrt{\pi} v ^2 _0 \Gamma (2 + \alpha / 2)}{4 (\alpha + 5) \Gamma (5/2 + \alpha / 2)}.
\end{equation}
\end{enumerate}
Spatial distance are in units of the effective radius of Draco, and velocities, in effective radii per $10^7$ years.

The role of the M2M method is to determine the value of $\alpha$ which best fits the data, and then we use the circular velocity to determine the mass it implies (via equations \ref{eqn:wilkcirc} and \ref{eqn:gencirc}) and thus the mass-to-light ratio.  We vary $\alpha$ in the same manner that the mass-to-light $\Upsilon$ ratio was varied in section \ref{sec:appml}.

\subsection{Proof of approach}\label{sec:dracoconcept}
Before using Kleyna's Draco data, we examine the ability of the M2M method to determine the Wilkinson $\alpha$ parameter.  Surface brightness and line-of-sight second velocity moment data are created using the functions in section \ref{sec:dracointro} with $1\%$ relative errors for surface brightness and $10\%$ for the velocity moment. For this trial data set, $\alpha = -0.5$. We run the M2M models with $2 \times 10^5$ particles and for $250$ time units.  The model parameters have values $\epsilon = 2.5 \times 10 ^{-3}$, $\mu = 1$, $\lambda _{\rmn{sum}} = 10^3$ and $\lambda _{\rmn{iso}} = 7 \times 10^{-2}$. $\lambda _{\rmn{SB}}$ has the fixed value $\lambda _{\rmn{SB}} = 10 ^{-5}$ while $\lambda _{\rmn{VD}}$ is varied to accommodate the different maximum velocities in the models as $\alpha$ is varied (from $\alpha = -0.8$ to $\alpha = -0.2$).  We find that the modelling runs give clear minima (close to the true $\alpha$ value) in the $\chi^2$ values, with $\alpha = -0.51$ from $\chi ^2 _{\rmn{SB}}$ and $\alpha = -0.49$ from $\chi ^2 _{\rmn{VD}}$.  That is, the approach works !

\subsection{Modelling Draco}\label{sec:modeldraco}
\begin{figure}
		\centering
		\includegraphics[width=84mm]{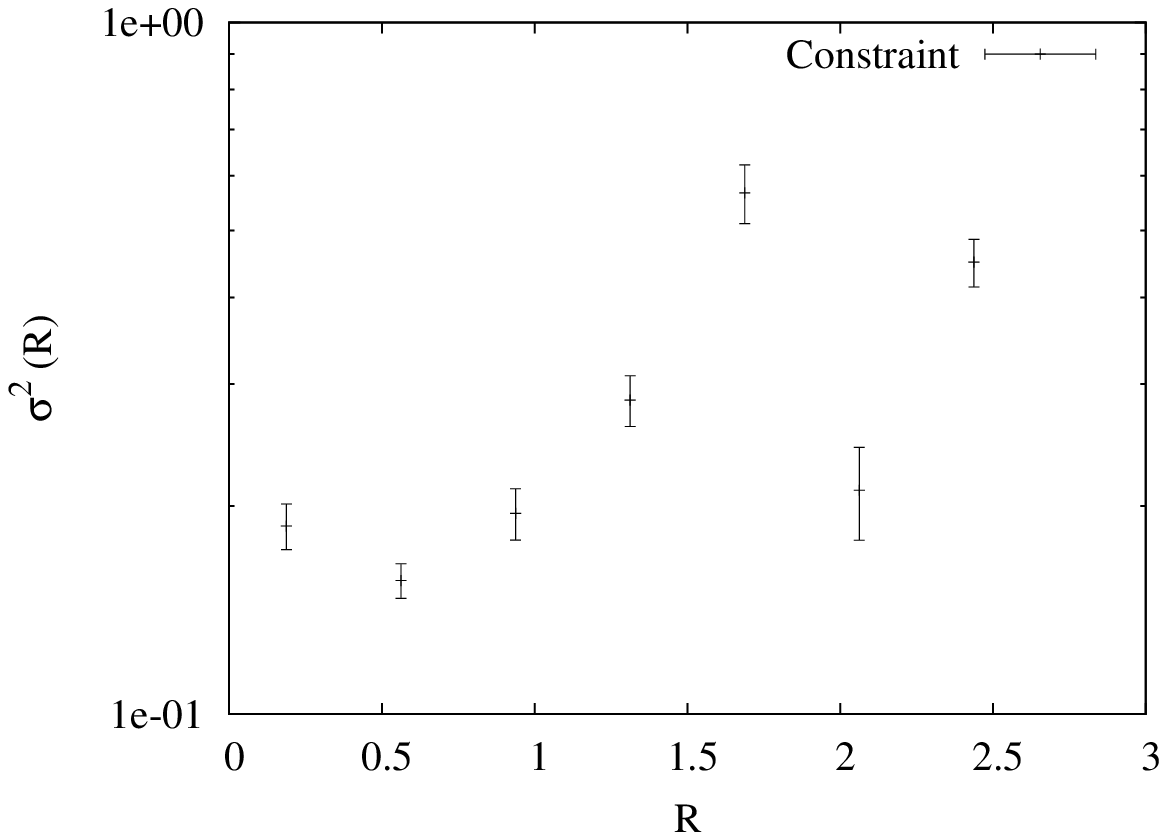}
		\includegraphics[width=84mm]{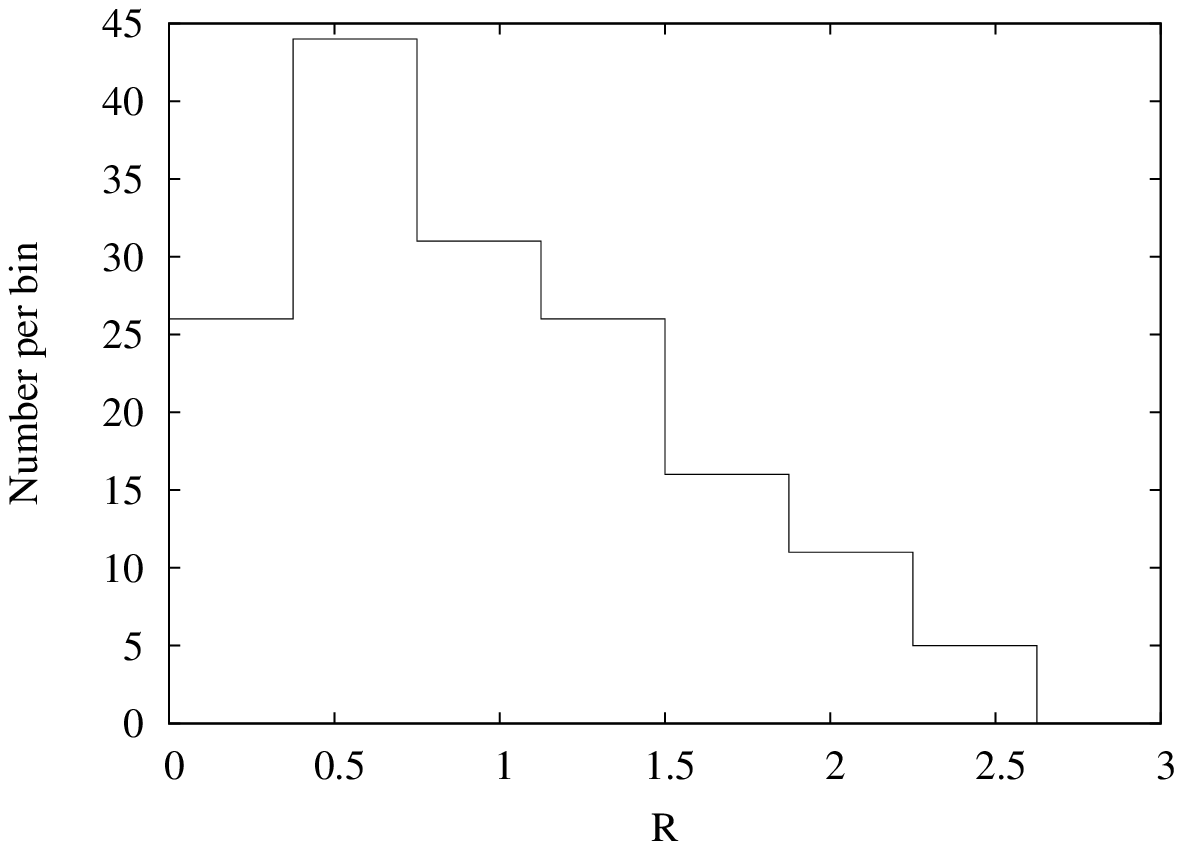}
		\caption[Draco second velocity moment]{Draco second velocity moment from binned velocity measurements.  The top panel shows the second velocity moment data points and error terms, and the bottom panel, the number of velocity measurements per bin. The projected radius $R$ is in units of the effective radius of Draco, and velocities are in effective radii per $10^7$ years.}
	\label{fig:bindracodata}
\end{figure}

\begin{figure}
		\centering
		\includegraphics[width=84mm]{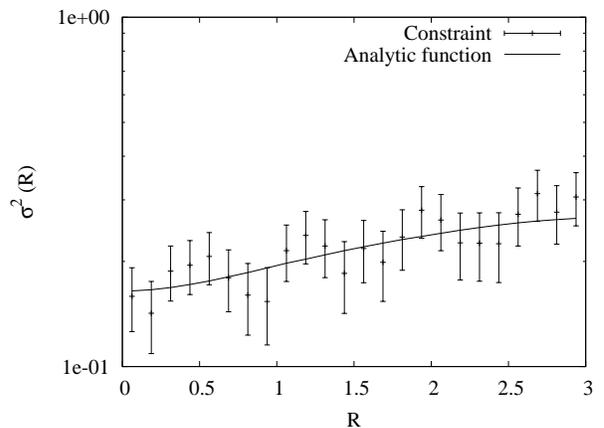}
		\caption[Trial second velocity moment data]{The trial second velocity moment data used in the proof of concept exercise in section \ref{sec:dracoconcept}.}
	\label{fig:bintrialdata}
\end{figure}

Binning the Draco velocity measurements (159 in total), using equal interval projected radius bins, to create a set of second velocity moment data points and error terms gives the plots in Figure \ref{fig:bindracodata}.  As can be seen, the measurements are centrally clustered and the velocity moment data points at higher radii suffer from a lack of contributing measurements.
Comparing Figures \ref{fig:bindracodata} and \ref{fig:bintrialdata}, the trial data has more data points than the Draco data ($24$ versus $7$) and the spread of points reflects the generating function.  The Draco data does not visibly reflect any underlying curve.

After some experimentation, we run the M2M models for values of the Wilkinson $\alpha$ parameter in the range $-1.5$ to $-0.5$ with  the same model parameters as in the proof of concept exercise in section \ref{sec:dracoconcept}.  The results are captured in Table \ref{tab:dracoresults}.  
\begin{table}
	\begin{center}
	\caption{M2M determination of Draco potential power law}
	\medskip
	\label{tab:dracoresults}
	\begin{tabular}{ccc}
		\hline
		\textbf{Measure} & $\bm{\alpha}$ & \textbf{Error bounds} \\
		\hline
		$-F$ & $-1.18$ & \\
		$\chi ^2 _{\rmn{LM}}$ & $-1.24$ & $+0.12$, $-0.16$ \\
		$-S$ & $-0.96$ & \\
		$\chi ^2$ & $-0.95$ & $+0.04$, $-0.06$ \\
		$\chi ^2 _{\rmn{SB}}$ & $-0.96$ & $+0.03$, $-0.04$ \\
		$\chi ^2 _{\rmn{VD}}$ & $-0.90$ & $+0.36$, $-0.35$ \\
		\hline
	\end{tabular}
	\end{center}
\medskip
The value of the potential power law index $\alpha$ determined from different M2M model outputs.
\end{table}
Weight convergence is high ($> 99\%$) in all runs except for $\alpha = -1.5$ where it is slightly lower ($96\%$). There are no issues associated with the sum of weights constraint and the isotropic dispersion constraint.

We repeat the runs and include the individual velocity measurements as constraints (see section \ref{sec:theory}, equation \ref{eqn:indivobs}) as well as the line-of-sight second moment. The individual measurements make very little difference to the results in Table \ref{tab:dracoresults}.  Taking the model produced line-of sight velocity distribution and comparing it to the theoretical distribution at all of the individual velocity measurement points gives a measure of how well the M2M model has reproduced the theoretical distribution (Table \ref{tab:dracolosvd}).
\begin{table}
	\begin{center}
	\caption{Model vs theory line-of-sight velocity distribution}
	\medskip
	\label{tab:dracolosvd}
	\begin{tabular}{ccc}
		\hline
		$\bm{\alpha}$ & \textbf{\% within} $\mathbf{5\%}$ & \textbf{\% within} $\mathbf{10\%}$ \\
		\hline
		$-1.50$ & $66.0$ & $88.7$ \\
		$-1.25$ & $84.3$ & $100.0$ \\
		$-1.10$ & $89.3$ & $100.0$ \\
		$-1.00$ & $93.7$ & $100.0$ \\
		$-0.90$ & $99.4$ & $100.0$ \\
		$-0.75$ & $88.1$ & $98.1$ \\
		$-0.50$ & $81.8$ & $90.6$ \\
		\hline
	\end{tabular}
	\end{center}
\medskip
For different values of $\alpha$, the percentage of the individual velocity measurements that have model line-of-sight velocity distribution probabilities within $5\%$ and $10\%$ of their theoretical probabilities.
\end{table}
As can be seen the highest reproduction at the $5\%$ level occurs at $\alpha = -0.9$ which is also the value resulting for $\chi ^2 _{\rmn{VD}}$ in Table \ref{tab:dracoresults}.  

\begin{figure}
	\centering
	\includegraphics[width=84mm]{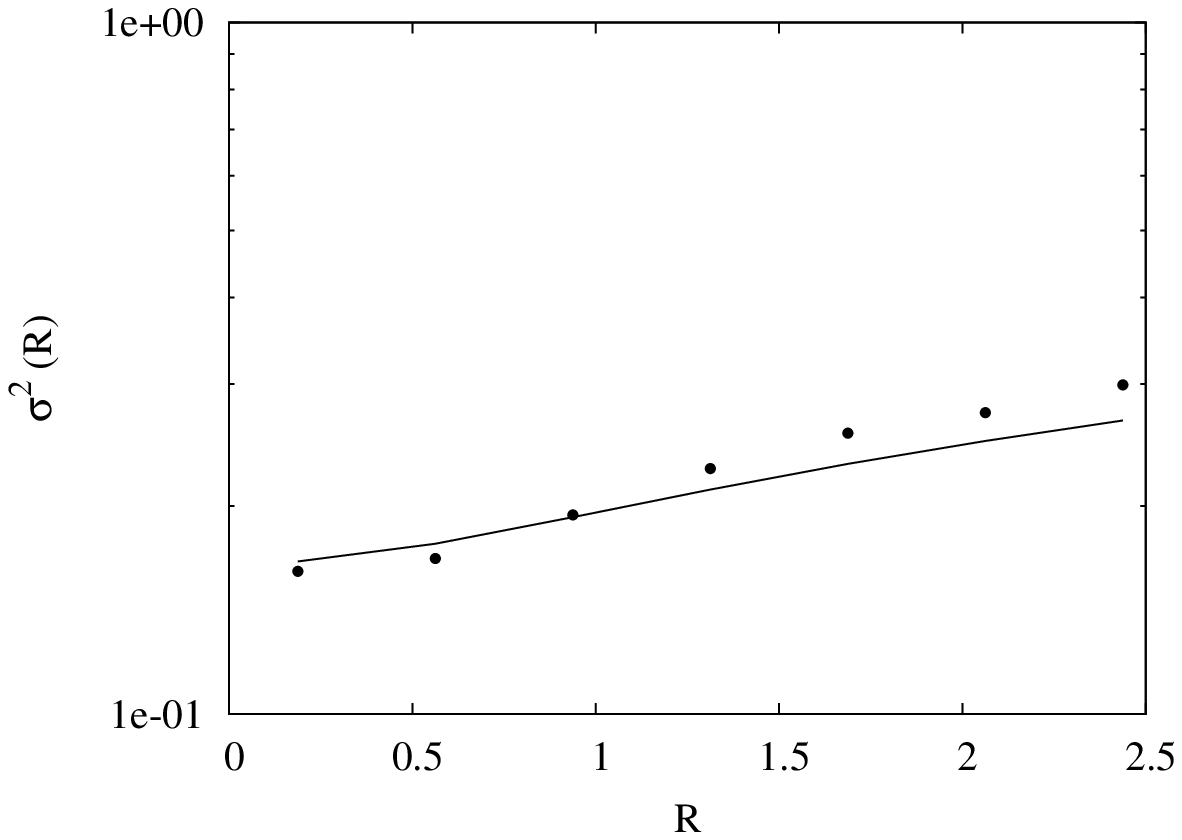}
	\includegraphics[width=84mm]{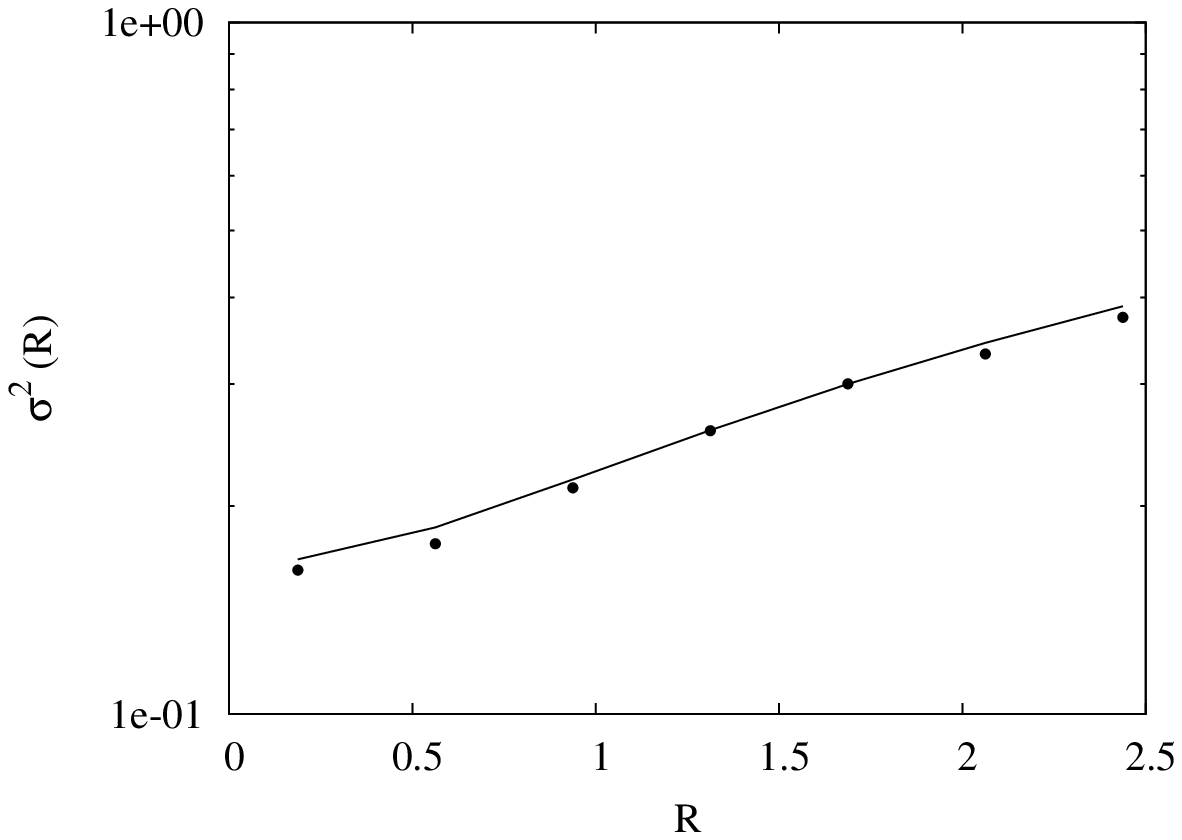}
	\includegraphics[width=84mm]{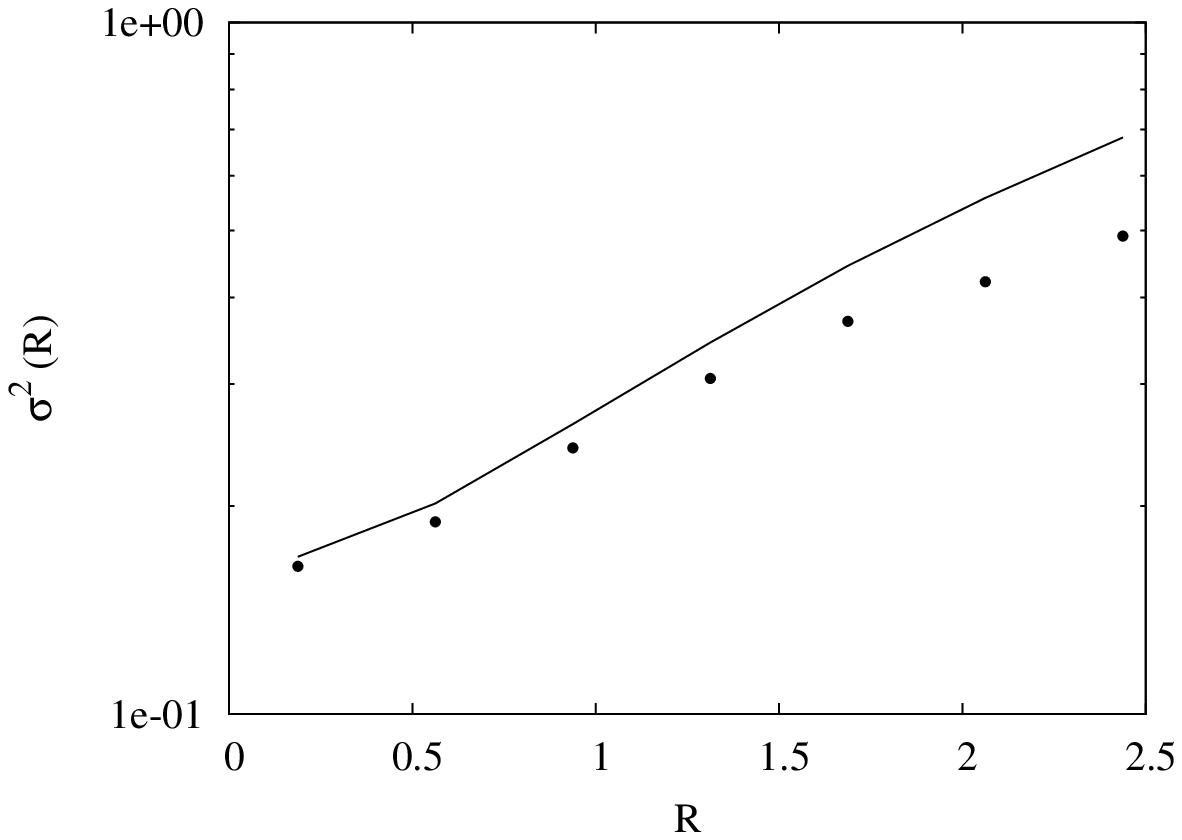}\\
	\caption[Draco second velocity moment reproduction]{Comparison of the model produced line-of-sight second velocity moment (solid points) with the theoretical moment for $\alpha = -0.5$ (top panel), $\alpha = -0.9$ (middle panel), and $\alpha = -1.5$ (bottom panel).}
	\label{fig:dracoreprovd}
\end{figure}

Comparing how well the model line-of-sight second velocity moment matches the theoretical moment for a given value of $\alpha$ (Figure \ref{fig:dracoreprovd}) shows that a good match is achieved for $\alpha = -0.9$.  For $\alpha = -0.5$, the model overestimates the theoretical moment implying that the magnitude of $\alpha$ needs to be increased, and for $\alpha = -1.5$ underestimates it implying that a reduction in the magnitude of $\alpha$ is required.

Using $\alpha = -0.90^{+0.36} _{-0.35}$, the mass within 3 core radii is $(9.7\pm 2.3) \times 10^7 \ \rmn{M}_{\odot}$.  \citet{Kleyna2002} achieved a value of $8 ^{+3} _{-2} \times 10^7 \ \rmn{M}_{\odot}$.  Taking Draco's V-band luminosity as $(1.8 \pm 0.8) \times 10^5 \  \rmn{L}_{\odot}$  \citep{Irwin1995}, the mass-to-light ratio for Draco is $539\pm 136 \ \rmn{M}_{\odot}/\rmn{L}_{\odot}$.  \citet{Kleyna2002} obtained a lower value of $440\pm 240 \ \rmn{M}_{\odot}/\rmn{L}_{\odot}$ which is not surprising given they were using an anisotropic dispersion model.

\section{Conclusion}
Other authors have commented on the potential of Syer \& Tremaine's made-to-measure-method.  We hope in this paper that we have helped to expose some of that potential in a practical way.  We believe we have added clarity to the construction of the weight evolution equation, shown how constraints may be incorporated and defined 2 further constraints. We have demonstrated a simple application of M2M modelling by determining the mass-to-light ratio of a theoretical Plummer model.  With a variety of constraints and initial conditions, we arrive in all cases at a model value close to the true value.   Using the method with an isotropic velocity dispersion model,  we estimate the mass-to-light ratio of Draco and achieve a V-band value of  $539\pm 136 \ \rmn{M}_{\odot}/\rmn{L}_{\odot}$.

We encourage others to use the method - particularly, for comparison purposes, those who have experience of using Schwarzschild's method. It should be noted that much of the preparatory work (whether theoretical, or manipulation of observational data, say) leading to the execution of a Schwarzschild or a M2M model is in fact common.  The key difference in the methods is how the final particle weights are determined. Also, the orbit selection stage in a Schwarzschild model is not required for a M2M model.  The only shortfall against Schwarzschild's method we are aware of is the modelling of proper motion data.

We have given insight into how a made-to-measure model may be implemented and provided some information on its behaviour and how to size and tune such a model.  We have defined a simple mechanism for assessing the degree of particle weight convergence and are not aware of such a mechanism being used elsewhere with M2M models.  From a software perspective, our first unparallelised implementation was less than $2000$ lines of C code so the effort required to establish a working prototype is not huge.

As this paper was being completed, \citet{Dehnen2009} became available and there is some overlap with this paper notably in total weight conservation and the use of the $\beta$ anisotropy parameter as a constraint.

Our next steps are to continue the move away from using theoretical models and apply the method to observations of more, real galaxies, and to understand practically the relative strengths of Schwarzschild's method and the made-to-measure method.

\section*{Acknowledgements}
RJL acknowledges receipt of an STFC postgraduate studentship.  The authors would like to express their thanks to Scott Tremaine, James Binney and John Magorrian for various fruitful discussions, and to Wyn Evans for advice and guidance throughout.

\bibliographystyle{mn2e}
\bibliography{rjlmodel}

\label{lastpage}
\end{document}